\documentclass{emulateapj}
\usepackage{natbib}
\usepackage{graphicx}
\usepackage{amsmath}
\usepackage{float}
\usepackage{booktabs}
\usepackage{afterpage}
\usepackage{flafter}
\newcommand{\nh} {$N_{\text{H}}$}
\newcommand{\chisq} {$\chi^{2}$}

\newcommand{\nustar} {\textit{NuSTAR}}
\newcommand{\xmm} {\textit{XMM-Newton}}
\newcommand{\xmms}{\textit{XMM-Newton}}
\newcommand {\msun} {M$_{\odot}$}
\newcommand{\degree} {$^\circ$}

\shorttitle{Modeling disk precession in LMC X-4 and SMC X-1}
\shortauthors{Brumback et al.}
\slugcomment{Accepted to The Astrophysical Journal}

\begin{document}

\title{Modeling the precession of the warped inner accretion disk in the pulsars LMC X-4 and SMC X-1 with \textit{NuSTAR} and \textit{XMM-Newton}}

\author{McKinley C. Brumback{\altaffilmark{1}}, Ryan C. Hickox{\altaffilmark{1}}, Felix S. F\"urst{\altaffilmark{3}}, Katja Pottschmidt{\altaffilmark{4,5}}, John A. Tomsick{\altaffilmark{6}}, J\"orn Wilms{\altaffilmark{7}}  }

\altaffiltext{1} {Department of Physics \& Astronomy, Dartmouth College, 6127 Wilder Laboratory, Hanover, NH 03755, USA} 
\altaffiltext{7} {Dr. Karl Remeis-Sternwarte and Erlangen Centre for Astroparticle Physics, Sternwartstrasse 7, 96049 Bamberg, Germany} 
\altaffiltext{3} {European Space Astronomy Centre (ESA/ESAC), Operations Department, Villanueva de la Ca$\tilde{\text{n}}$ada Madrid, Spain} 
\altaffiltext{4} {CRESST, Department of Physics and Center for Space Science and Technology, UMBC, Baltimore, MD 210250, USA} 
\altaffiltext{5} {NASA Goddard Space Flight Center, Code 661, Greenbelt, MD 20771, USA} 
\altaffiltext{6} {Space Sciences Laboratory, University of California, Berkeley, 7 Gauss Way, Berkeley, CA 94720, USA}

\begin{abstract}

We present a broad-band X-ray study of the effect of superorbital periods on X-ray spectra and pulse profiles in the neutron star X-ray binaries LMC X-4 and SMC X-1. These two sources display periodic or quasi-periodic variations in luminosity on the order of tens of days which are known to be superorbital, and are attributed to warped, precessing accretion disks. Using joint \nustar\ and \xmm\ observations that span a complete superorbital cycle, we examine the broad-band spectra of these sources and find the shape to be well described by an absorbed power law with a soft blackbody component. Changes in spectral shape and pulse profile shape are periodic with superorbital period, as expected from a precessing disk. We perform X-ray tomography using the changes in pulse profiles to model the geometry and kinematics of the inner accretion disk. Our simple beam and inner disk geometric model indicates that the long term changes in soft pulse shape and phase are consistent with reprocessed emission from a precessing inner disk. 
\end{abstract}

\section{Introduction} \label{sec:int}
X-ray pulsars are rotating, highly magnetized neutron stars that accrete gas from a stellar companion via Roche lobe overflow or stellar outflows (e.g.\ \citealt{nagase2001}). The gas falls gravitationally in a disk toward the pulsar until it reaches the magnetosphere, where magnetic pressure exceeds the ram pressure from the disk and forces accretion along the neutron star's dipole field onto the magnetic poles. If the magnetic axis is misaligned from the rotation axis, the accretion columns will rotate with the neutron star and cause bright beams of X-ray radiation to sweep across space. While this general picture is widely accepted, the fundamentals of magnetically-dominated accretion are still unclear. Understanding accretion within magnetic fields is essential to the study of accreting white dwarfs and young stellar objects, as well as X-ray pulsars. Magnetohydrodynamic simulations of accreting neutron stars have begun to explore the possible structures of magnetized accretion flows around neutron stars (e.g.\ \citealt{romanova2002,romanova2003,romanova2004}), however observational constraints on these flows are needed to fully develop our understanding of magnetically dominated accretion.

X-ray pulsars that display periodic or quasi-periodic superorbital variability are unique systems in which to observationally probe the nature of magnetically-dominated accretion and the structure of the inner accretion disk. Superorbital variabilities in some X-ray pulsars such as LMC X-4, SMC X-1, and Her X-1 are attributed to warped inner accretion disks that precess around the pulsar, causing fluctuations in luminosity on the order of tens of days (e.g.\ \citealt{gerend1976,heemskerk1989,wojdowski1998}). As the pulsar rotates, the neutron star's beam irradiates the warped disk, which reprocesses this emission into softer X-rays (\citealt{hickoxvrtilek2005}, hereafter HV05). The reprocessed pulsations exhibit a different pulse shape and phase from the hard X-ray emission, which is dominated by the pulsar beam (e.g.\ \citealt{neilsen2004,zane2004}). HV05 developed an irradiated warped disk model that used differences between the hard and soft pulsations to constrain the beam and disk geometry. However, the model requires both hard and soft X-ray coverage to fully constrain emission from both the pulsar beam and the disk.

\begin{figure*}
\plottwo{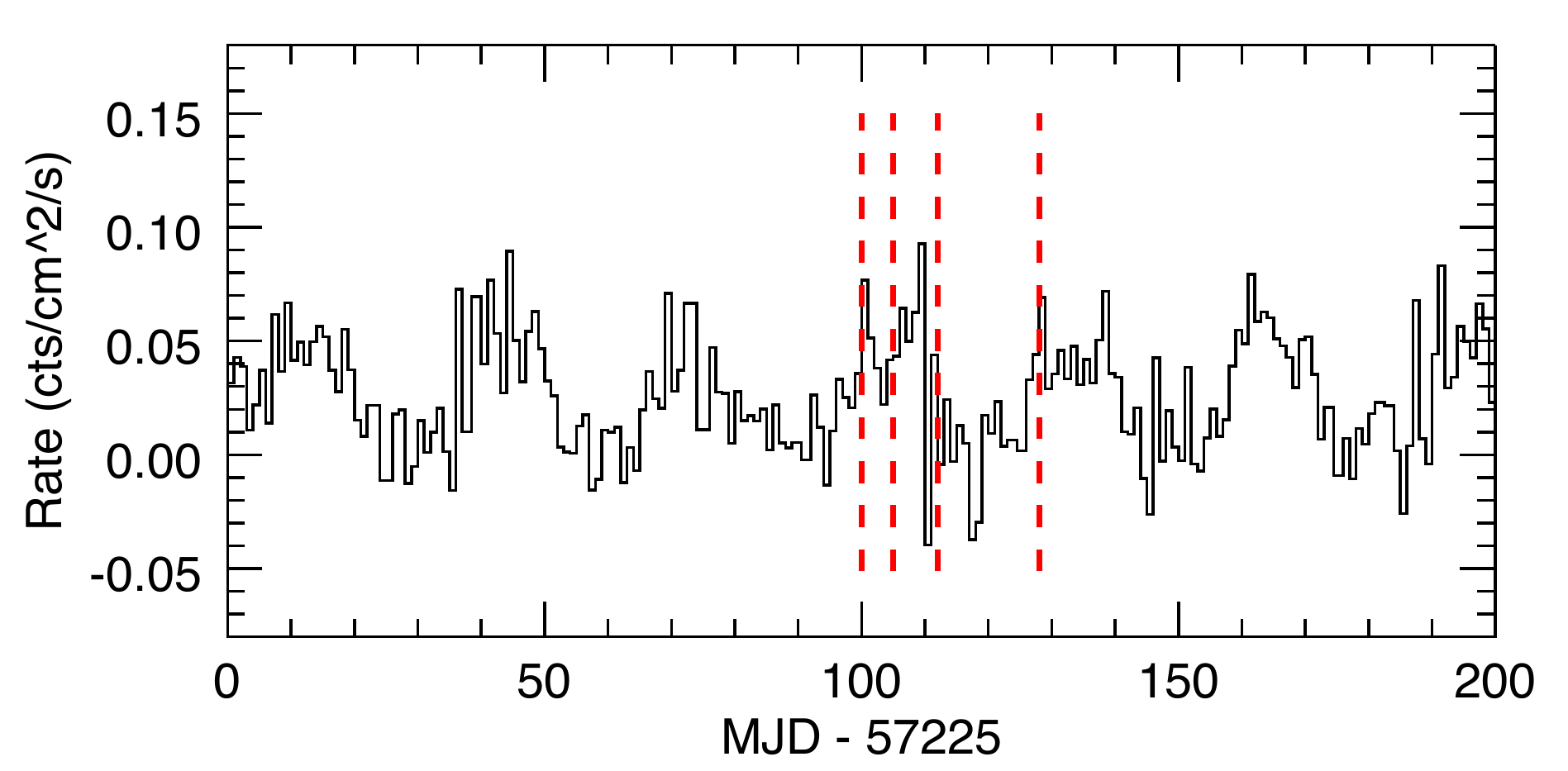}{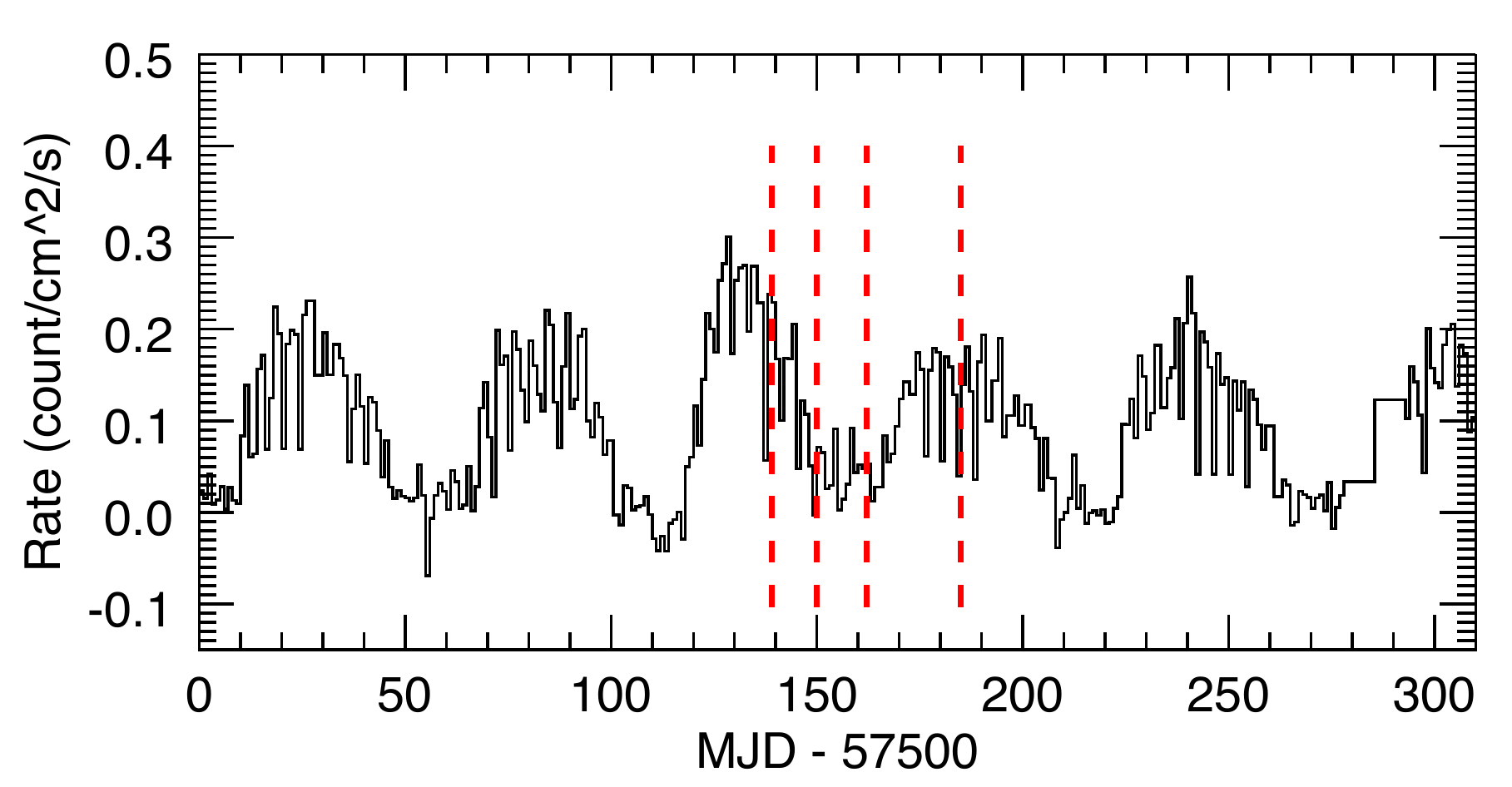}
\caption{One day averaged 2--20 keV MAXI (\citealt{matsuoka2009}) light curves of LMC X-4 (left) and SMC X-1 (right) during the period of observations. The red dashed lines mark the times of joint \xmm\ and \nustar\ observations.}
\label{fig:solc}
\end{figure*}

\begin{deluxetable*} {cccccc} 
\tablecolumns{6}
\tablecaption{Description of LMC X-4 Observations}  
\tablewidth{0pt}
\tablehead{
\colhead{Date} & \colhead{$\phi_{SO}$} & \colhead{Observation ID} & \colhead{Observatory} & \colhead{Telescope Mode}
& \colhead{Exposure (ks)}  }
\startdata
30 Oct.\ 2015 & 0.0 & 30102041002 & \nustar\ & \nodata & 24.6 \\
30 Oct.\ 2015 & 0.0 & 0771180101 & \xmm\ & Small Window & 20.7 \\
04 Nov.\ 2015 & 0.17 & 30102041004 & \nustar\ & \nodata & 21.9 \\
04 Nov.\ 2015 & 0.17 &  0771180201& \xmm\ & Small Window & 19.7 \\
11 Nov.\ 2015 & 0.4 & 30102041006 & \nustar\ & \nodata & 23.0 \\
11 Nov.\ 2015 & 0.4 & 0771180301 & \xmm\ & Small Window & 22.9 \\
27 Nov.\ 2015 & 1.0 & 30102041008 & \nustar\ & \nodata & 20.3 \\
27 Nov.\ 2015 & 1.0 & 0771180401 & \xmm\ & Small Window & 20.1
\enddata
\label{tab:lmcdat}
\end{deluxetable*}

\begin{deluxetable*} {cccccc} 
\tablecolumns{6}
\tablecaption{Description of SMC X-1 Observations}  
\tablewidth{0pt}
\tablehead{
\colhead{Date} & \colhead{$\phi_{SO}$} & \colhead{Observation ID} & \colhead{Observatory} & \colhead{Telescope Mode}
& \colhead{Exposure (ks)}  }
\startdata
8 Sept.\ 2016 & 0.1 & 30202004002 & \nustar\ & \nodata & 22.5 \\
8 Sept.\ 2016 & 0.1 & 0784570201 & \xmm\ & Fast Timing Mode & 20.9 \\
19 Sept.\ 2016 & 0.3 & 30202004004 & \nustar\ & \nodata & 21.1 \\
19 Sept.\ 2016 & 0.3 &  0784570301& \xmm\ & Fast Timing Mode & 20.9 \\
1 Oct.\ 2016 & 0.5 & 30202004006 & \nustar\ & \nodata & 20.4 \\
1 Oct.\ 2016 & 0.5 & 0784570401 & \xmm\ & Fast Timing Mode & 22.9 \\
24 Oct.\ 2016 & 1.1 & 30202004008 & \nustar\ & \nodata & 20.8 \\
24 Oct.\ 2016 & 1.1 & 0784570501 & \xmm\ & Fast Timing Mode & 22.9
\enddata
\label{tab:smcdat}
\end{deluxetable*}

The current era of X-ray astronomy offers a new opportunity to apply the HV05 warped disk model with sensitive hard X-ray coverage thanks to \nustar, which can constrain X-rays between 3 and 79 keV (\citealt{harrison2013}). In this paper, we use the hard X-ray sensitivity of \nustar\ combined with the soft X-ray coverage of \xmm\ to analyze the spectral and geometrical changes associated with disk precession in LMC X-4 and SMC X-1 within a single disk precession cycle. 

Previous works have examined the effect of superorbital cycle on pulse shape in LMC X-4 and SMC X-1, however these analyses lacked either the hard X-ray sensitivity necessary to constrain the pulsar beam or complete coverage of a single superorbital cycle. \cite{hung2010} used \textit{Suzaku} to observe LMC X-4 three times during the superorbital high state, however due to scheduling constraints these observations did not occur within a single superorbital cycle and thus cannot prove that changes in pulse profile shape are periodic with respect to superorbital phase. Additionally, the \textit{Suzaku} observations used in this work had poorer high energy sensitivity compared to \textit{NuSTAR}. \cite{neilsen2004} and HV05 used \textit{Chandra} and \xmm\ to observe changes in pulse profile shape in SMC X-1, however these observations did not occur within a single superorbital cycle and used \textit{XMM-Newton} and \textit{Chandra}, which did not probe above 10 keV. We therefore present the first broad-band X-ray observations of LMC X-4 and SMC X-1 spanning a complete superorbital cycle and re-sampling the first phase. 

LMC X-4 is a high mass X-ray binary in the Large Magellanic Cloud first detected by UHURU (\citealt{giacconi1972}). The binary contains a $1.57 \pm 0.11$ M$_\sun$ neutron star and its $18 \pm 1$ M$_\sun$ O8 giant companion (\citealt{kelley1983, falanga2015}). LMC X-4 is an eclipsing binary where the pulsar orbits its companion with a period of 1.4 d and rotates once every 13.5 s (\citealt{white1978}). Additionally, the binary has a superorbital period of 30.4 d (\citealt{lang1981, molkov2015}). LMC X-4 has a typical X-ray luminosity of $\sim$ 2 $\times 10^{38}$ erg s$^{-1}$, which is slightly less than the Eddington limit for neutron stars; however, this source exhibits frequent X-ray flares capable of reaching super-Eddington luminosities of a few $10^{39}$ erg s$^{-1}$ (e.g.\ \citealt{kelley1983,levine1991,moon2003,brumback2018b}).

SMC X-1 is an X-ray pulsar located in the Small Magellanic Cloud also discovered by UHURU (\citealt{leong1971}). The compact object is a 1.21 \msun\ (\citealt{falanga2015}) pulsar with a spin period of 0.7 s (\citealt{lucke1976}). The stellar companion is a B0 supergiant star and the binary orbit is 3.9 d (\citealt{schreier1972,webster1972, liller1973}). SMC X-1's superorbital period varies quasi-periodically between 40 to 60 days (\citealt{wojdowski1998, clarkson2003}). SMC X-1 is a bright binary, with a high state X-ray luminosity of $\sim$ 3 $\times 10^{38}$ erg s$^{-1}$.

In Section \ref{sec:data} of this work we will describe the joint \xmms\ and \nustar\ observations of LMC X-4 and SMC X-1 and their respective analysis procedures. We also describe the analysis of phase-averaged and phase-resolved spectroscopy and a timing analysis to extract pulse profiles. In Section \ref{sec:results} we introduce the HV05 warped disk model and use it to simulate our observed pulse profiles. We discuss the implication of these results in Section \ref{sec:disc}.

\section{Observations and Data Analysis} \label{sec:data}

\subsection{Observations}

\begin{figure*}
\plotone{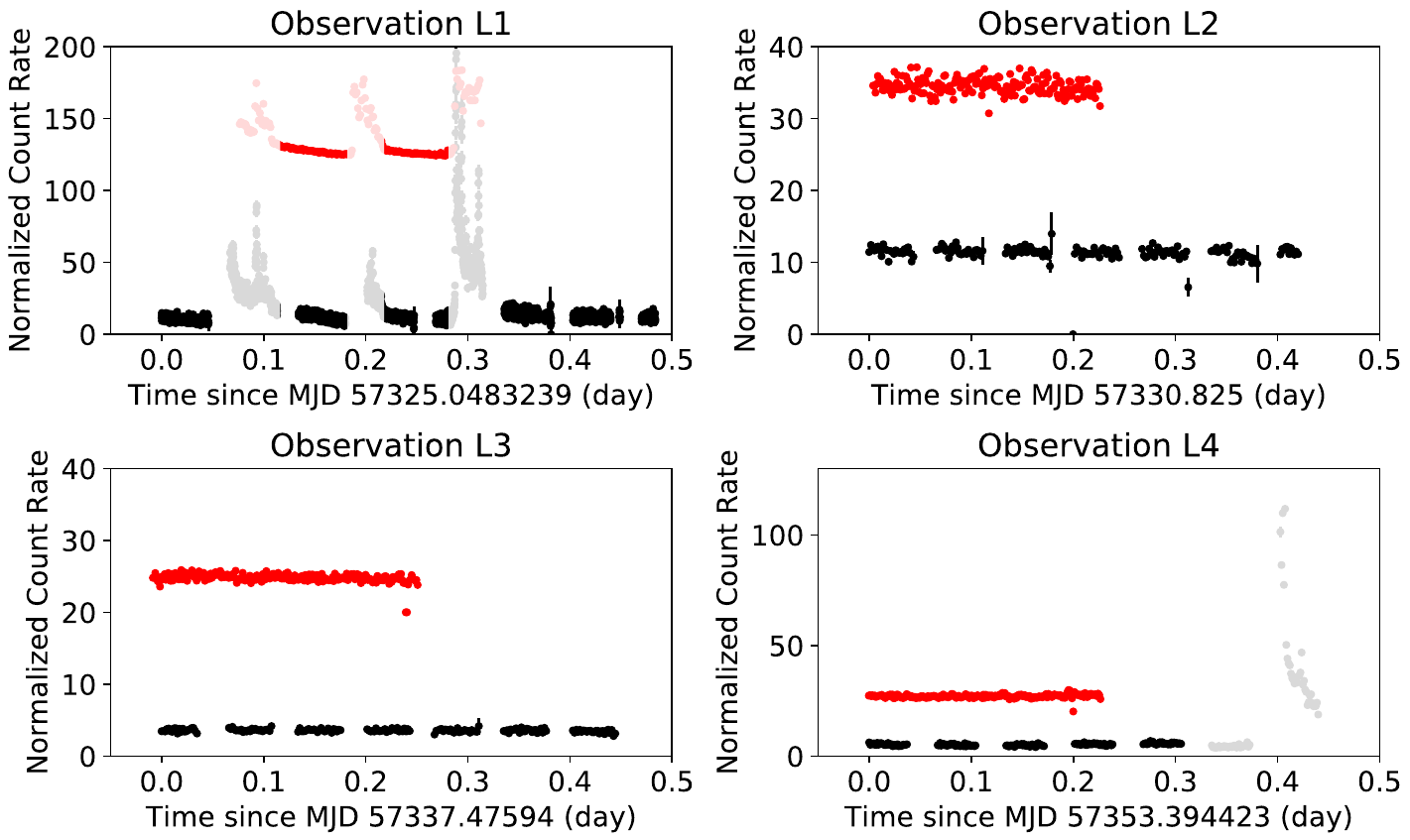}
\caption{\nustar\ 3--79 keV (black) and \xmm\ 0.2--12 keV (red) light curves of the four joint LMC X-4 observations. All \textit{XMM-Newton} light curves have been offset by a count rate of 20 for clarity, except for Observation L1 which has been offset by 120. We have excluded bright accretion flares from the light curves in Observations L1 and L4; excluded time intervals are shown with reduced opacity.}
\label{fig:lmclc}
\end{figure*}

\begin{figure*}
\plotone{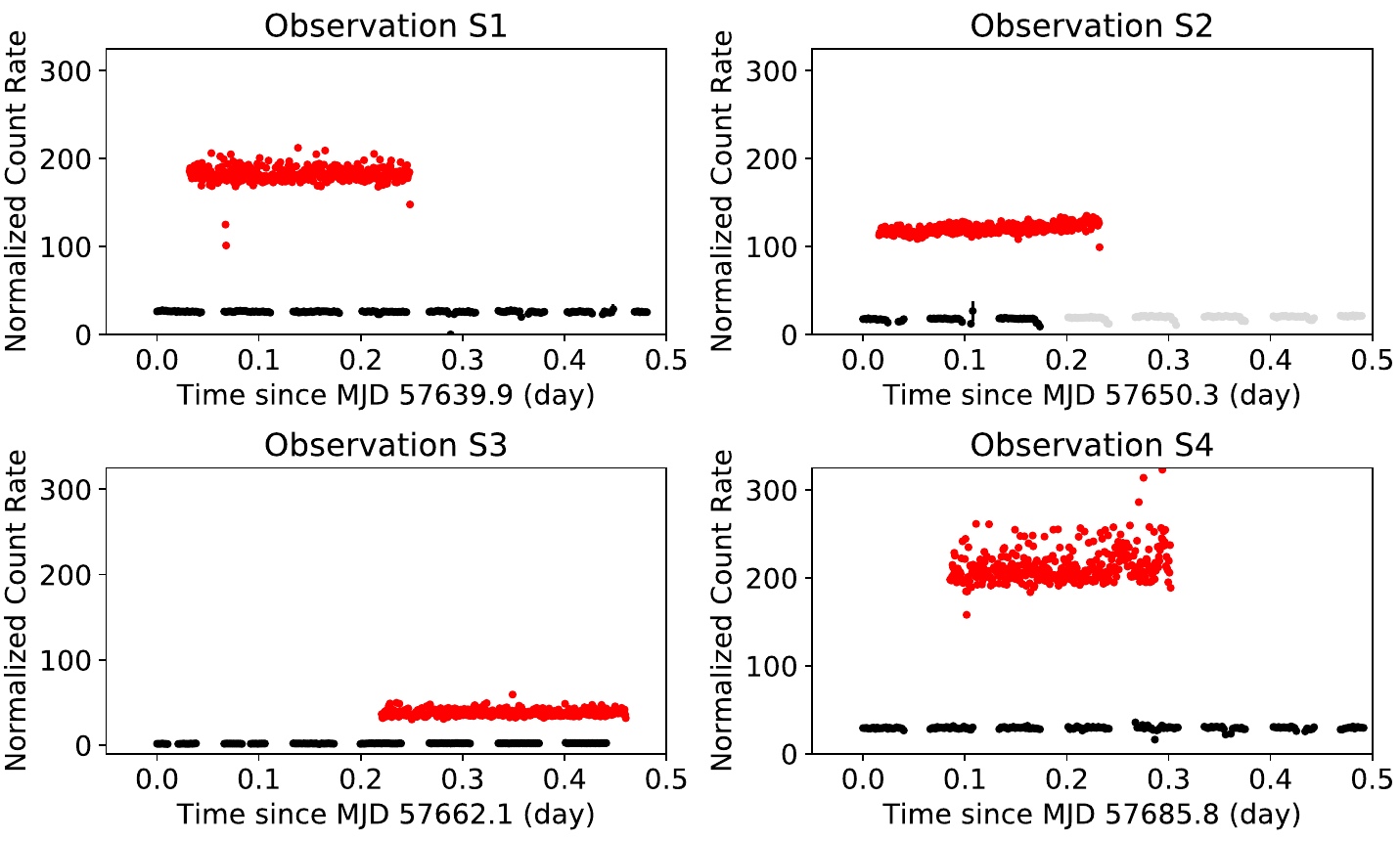}
\caption{\nustar\ 3--79 keV (black) and \xmm\ 0.2--12 keV (red) light curves of the four joint SMC X-1 observations. The y-axis has the same scale to show the low count rates in Observation S3. In Observation S3 only, the \textit{XMM} light curve has been offset by a count rate of 20 for clarity. In Observation S2, we only used the first part of the \nustar\ observation (black) in this analysis due to pulsation shape changes later in the observation. This pulsation change is energy dependent and only affects the \nustar\ pulse profile, so therefore we use the full \xmm\ observation. The full \nustar\ observation is shown with reduced opacity.}
\label{fig:smclc}
\end{figure*}

The observations used in this analysis consist of two distinct data sets. \nustar\ and \xmm\ observed LMC X-4 jointly at four different epochs between 30 October 2015 and 27 November 2015. Table \ref{tab:lmcdat} lists the observation ID numbers, dates, and exposure times for the LMC X-4 observations. \nustar\ and \xmm\ also observed SMC X-1 jointly at four epochs between 8 September 2016 and 24 October 2016, and Table \ref{tab:smcdat} contains the observation information for these observations. Figure \ref{fig:solc} shows the one day averaged MAXI light curves for LMC X-4 and SMC X-1 during the time of observations and indicates the time of observations. For both sources, our observations sample a single superorbital phase.

\subsubsection{LMC X-4 Data Analysis}
We reduced the \nustar\ data for LMC X-4 and SMC X-1 using version 1.8.0 of the NuSTARDAS pipeline and CALDB v20170727. For the \xmm\ data, we used version 14.0.0 of XMMSAS, with an updated leap second data file.

For each \nustar\ observation, we used DS9 to select circular source regions with a radius of 120 arcseconds centered on the source coordinates. A background region of the same size was selected away from the source. The \xmm\ observations were taken with EPIC-pn in Small Window Mode to minimize pile up, and we exclusively used the EPIC-pn instrument and not EPIC-MOS for the best timing resolution. In these observations, the source was positioned close to the edge of the chip and we detected small amounts of pile up using the XMMSAS tool  {\fontfamily{qcr}\selectfont epatplot}. To mitigate both of these effects, we selected annular source regions with an inner radius set to minimize pileup and the outer radius remaining on the chip. We found typical inner and outer radii for the annular source regions of 13 arcseconds and 42 arcseconds, respectively. We selected \xmm\ background regions from circular regions of radius 60 arcseconds, located away from the source area. We filtered all EPIC-pn data to contain only single and double events. We applied a barycentric correction to both the \nustar\ and \xmm\ data sets using the NuSTARDAS tool {\fontfamily{qcr}\selectfont barycorr} and the XMMSAS tool {\fontfamily{qcr}\selectfont barycen}, respectively. We also corrected the pulse arrival times using the LMC X-4 ephemeris described in \cite{levine2000}.

We show the light curves for the LMC X-4 observations in Figure \ref{fig:lmclc}, where the 0.2--12 keV \textit{XMM-Newton} light curves have been arbitrarily offset from the 3--79 keV \nustar\ light curves for clarity. Several bright accretion flares appear in observations L1 and L4. \cite{brumback2018b} found that these flares contain changes in pulse strength, shape, and phase, which could complicate the relative phase mapping presented in this analysis. For this reason, all flares have been removed from the light curve and only times of direct simultaneous observation between \xmms\ and \nustar\ have been used.

\subsubsection{SMC X-1 Data Analysis}
We used the same versions of NuSTARDAS and XMMSAS mentioned in the LMC X-4 data analysis to reduce the SMC X-1 data. For each \nustar\ observation, we used DS9 to select circular source regions of 120 arcseconds centered on the source coordinates. We selected background regions of the same size away from the source. For the \xmm\ observations, we used data from the EPIC-pn instrument in Timing Mode for the best timing resolution. We selected the source region from a column 20 pixels wide, centered on the source. The XMMSAS tool {\fontfamily{qcr}\selectfont epatplot} revealed slight amounts of pileup, and so we excised the brightest central pixel of the source to minimize this effect. We filtered the EPIC-pn data to contain only single and double events. We applied a barycentric correction to both data sets using the NuSTARDAS tool {\fontfamily{qcr}\selectfont barycorr} and the XMMSAS tool {\fontfamily{qcr}\selectfont barycen}, respectively. We also corrected for the pulse arrival times using the SMC X-1 ephemeris described in \cite{falanga2015}.

Figure \ref{fig:smclc} shows the \nustar\ 3--79 keV and \xmm\ 0.2--12 keV light curves for the SMC X-1 observations. The light curve from Observation S3 has a very low count rate, indicating that the source was weakly detected. As seen in Figure \ref{fig:solc}, the superorbital period sampled during our four SMC X-1 observations had a different amplitude and period from the preceding superorbital periods. The variable behavior of SMC X-1's superorbital phase has been previously monitored (e.g.\ \citealt{hu2011, hu2013, dage2019}) and could be caused by an instability in the accretion disk's radiation driven warp (\citealt{ogilvie2001}) The variation in superorbital phase occuring during our observations caused Observation S3 to occur during the low state of the superorbital cycle. We did not detect pulsations during this observation, and we therefore exclude it from further analysis. We also excluded part of the \nustar\ observation for Observation S2, which exhibited changes in pulse behavior not covered simultaneously by \xmm. This pulsation change is energy dependent and only affects the \nustar\ pulse profile, so therefore we use the full \xmm\ observation. The data used in this analysis is plotted in black in Figure \ref{fig:smclc}, and the full \nustar\ observation is shown with reduced opacity.

\subsection{Timing Analysis}

We used epoch folding, via the function {\fontfamily{qcr}\selectfont epfold} found in the Remeis observatory ISISscripts, to find the best period of each LMC X-4 observation, and a Monte Carlo simulation of 500 light curves to find the uncertainties. We used the \nustar\ data in the epoch folding analysis because it was more strongly pulsed than the \xmm\ data. The best pulse periods are listed in Table \ref{tab:lmcperiod}. 

\begin{figure*}
\centering
\begin{tabular}{cc}
	\includegraphics[scale=0.45]{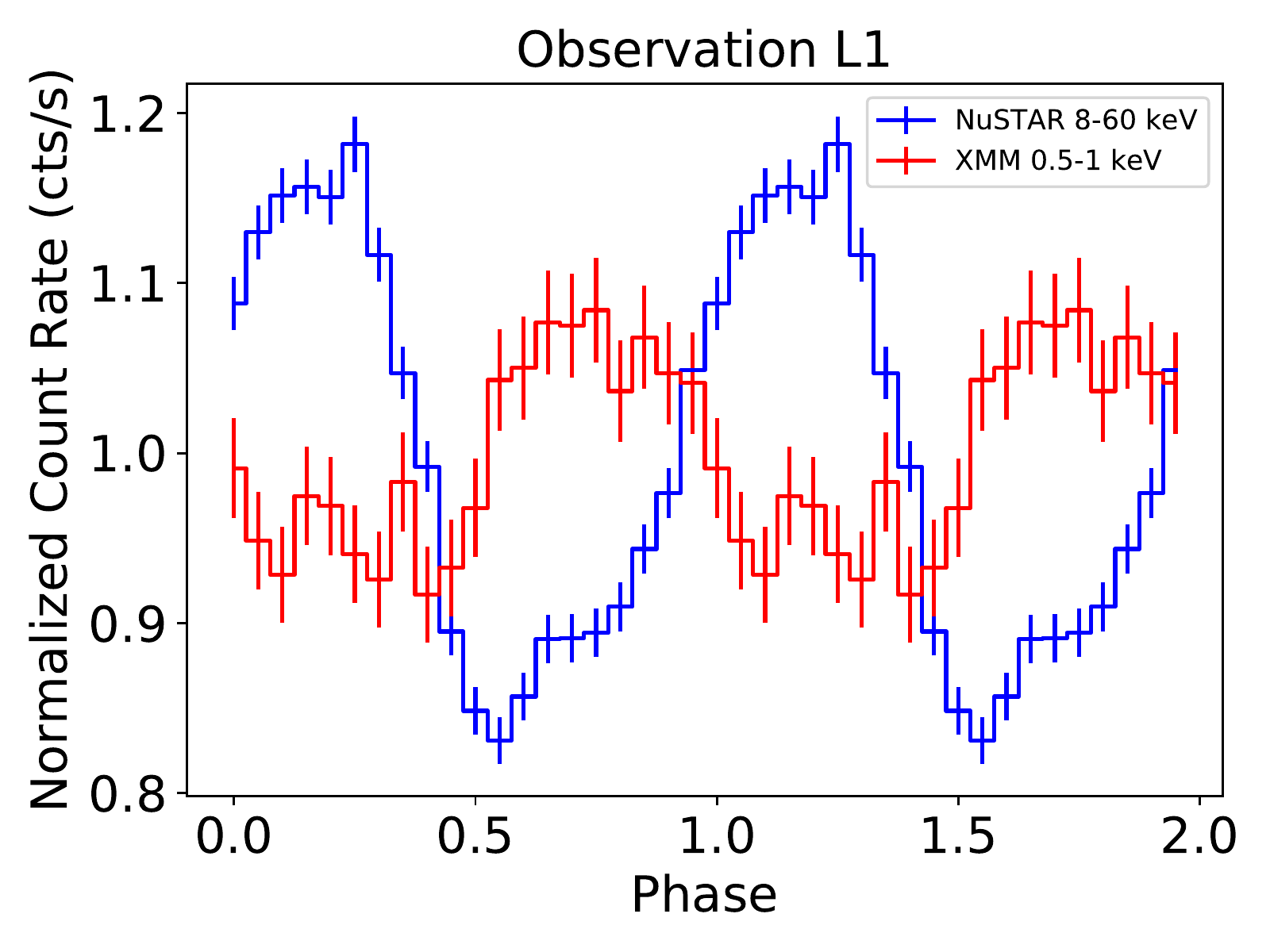} & \includegraphics[scale=0.45]{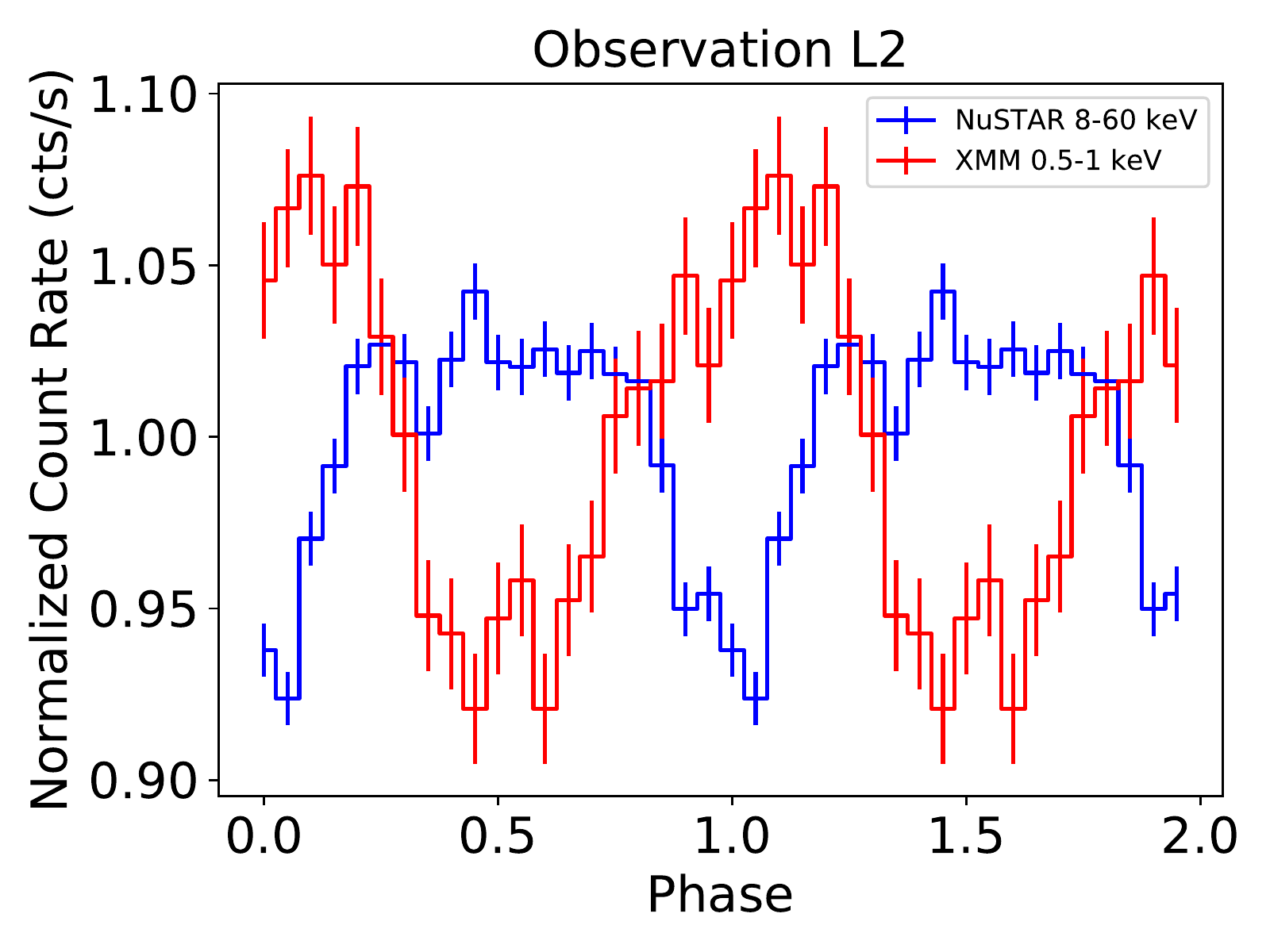} \\
	\includegraphics[scale=0.45]{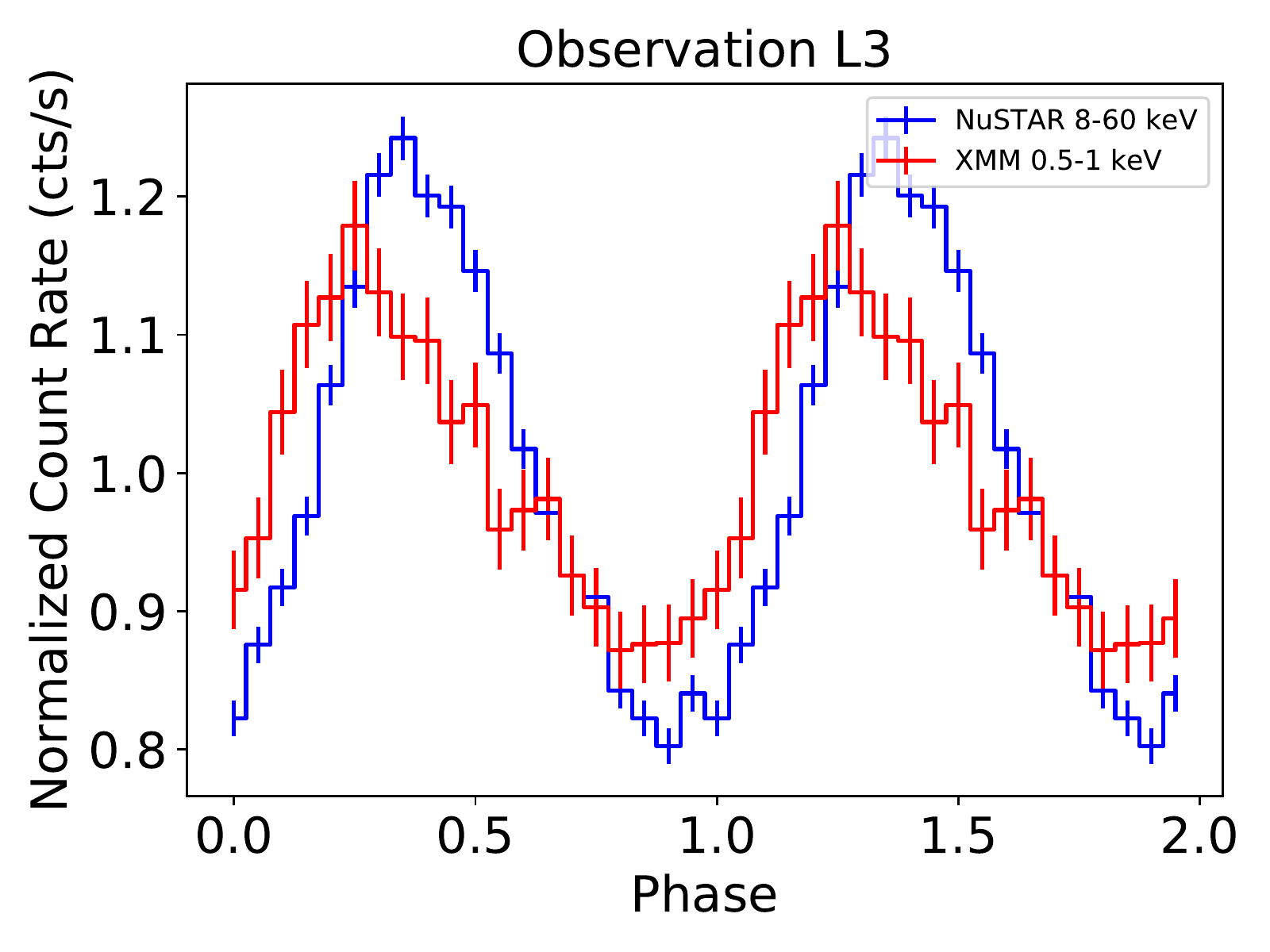} & \includegraphics[scale=0.45]{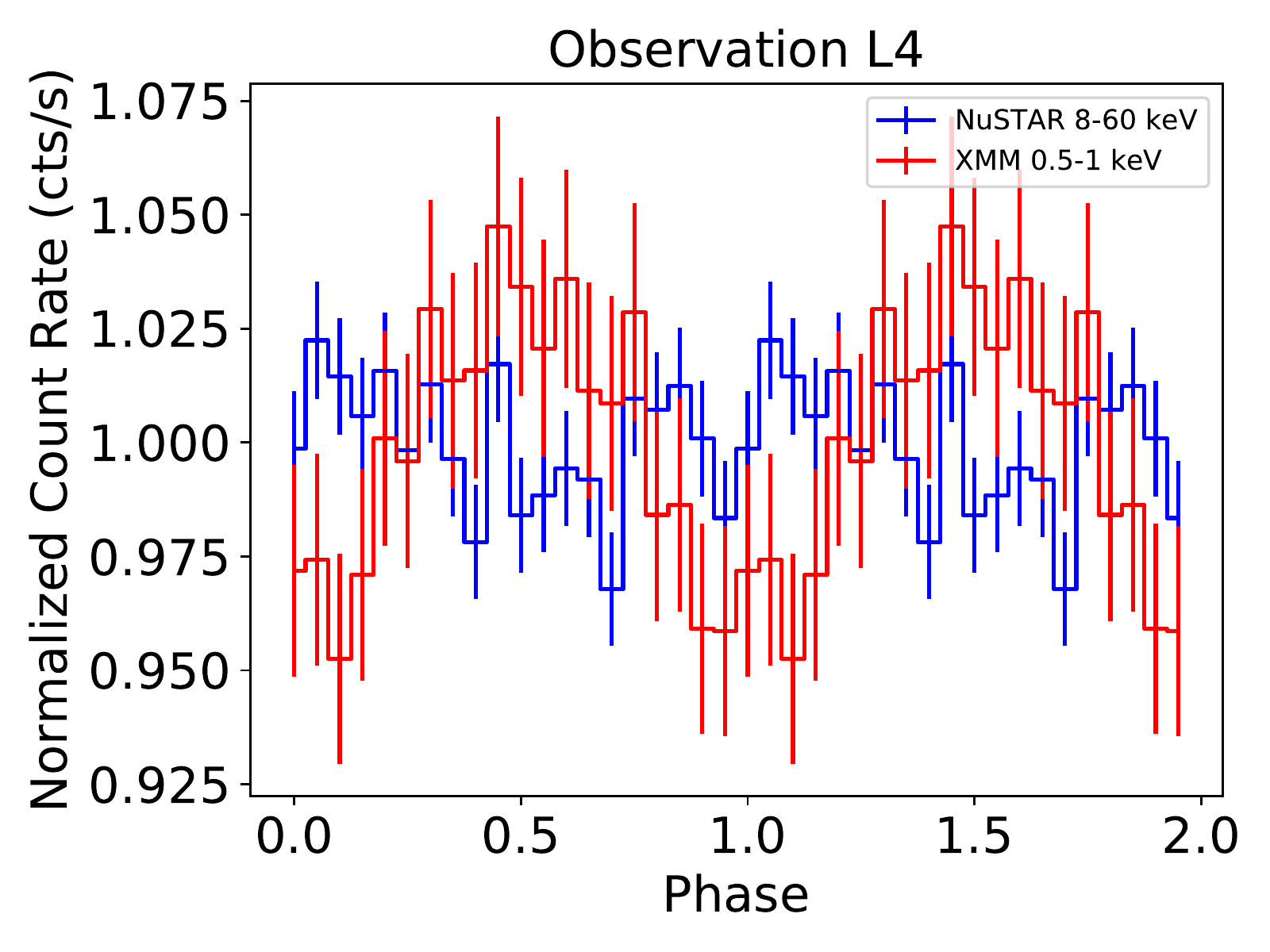}
\end{tabular}
\caption{Joint pulse profiles for the four LMC X-4 observations. Relative phase shifts are apparent between the \nustar\ 8--60 keV pulse profile (blue) and the \xmm\ 0.5--1 keV pulse profile (red). The change in relative phase from out of phase in Observations L1 and L2, to in phase in L3, and out of phase again in L4 is consistent with covering a complete precession cycle of the inner accretion disk.}
\label{fig:lmcpp}
\end{figure*}

Before creating pulse profiles, we filtered the event files by energy so that the \nustar\ data probed the hard pulsar beam emission (8--60 keV) and the \xmm\ data captured the soft reprocessed emission (0.5--1 keV). We created energy resolved pulse profiles using the folding technique in the FTOOL {\fontfamily{qcr}\selectfont efold}, which folds the light curve of each observation by the best period for that observation (see Figure \ref{fig:lmcpp}). We used 20 bins per phase for these pulse profiles.

\begin{deluxetable} {cc} 
\tablecolumns{2}
\tablecaption{Best Fit Spin Periods for LMC X-4 Observations}  
\tablewidth{0pt}
\tablehead{
\colhead{Observation} & \colhead{Spin Period (s)} }
\startdata
L1 & 13.5033 $\pm$ 0.0001 \\
L2 & 13.5028 $\pm$ 0.0001 \\
L3  & 13.50135 $\pm$ 0.0001 \\
L4 & 13.5003 $\pm$ 0.0009 
\enddata
\label{tab:lmcperiod}
\end{deluxetable}

We found that the Monte Carlo error analysis that we employed for LMC X-4 was not practical for determining uncertainties in the SMC X-1 spin period because of the timing resolution needed to evaluate the $\sim$0.7 s period. To make the analysis less computationally expensive, we employed the epoch folding technique found in the HENDRICS software (\citealt{hendrics}) tool {\fontfamily{qcr}\selectfont folding\_search}. This epoch folding tool searches the spin frequency and frequency first derivative simultaneously and returns a distribution of $Z^{2}_{4}$ statistics (\citealt{buccheri1983}). To estimate the 1$\sigma$ level uncertainty, we fitted this distribution 2-dimensional Gaussian using the Astropy model {\fontfamily{qcr}\selectfont Gaussian2D} and a Levenberg-Marquardt least squares fitting routine.

We confirmed that this epoch folding analysis is consistent with the Monte Carlo analysis from LMC X-4 by using {\fontfamily{qcr}\selectfont folding\_search} on LMC X-4 observations with high signal to noise. We found the results from each method to be consistent, and therefore do not believe that the difference in method will affect our measured pulse periods. The best pulse periods for the SMC X-1 data are listed in Table \ref{tab:smcperiod}.

\begin{figure*}
\centering
\begin{tabular}{cc}
	\includegraphics[scale=0.45]{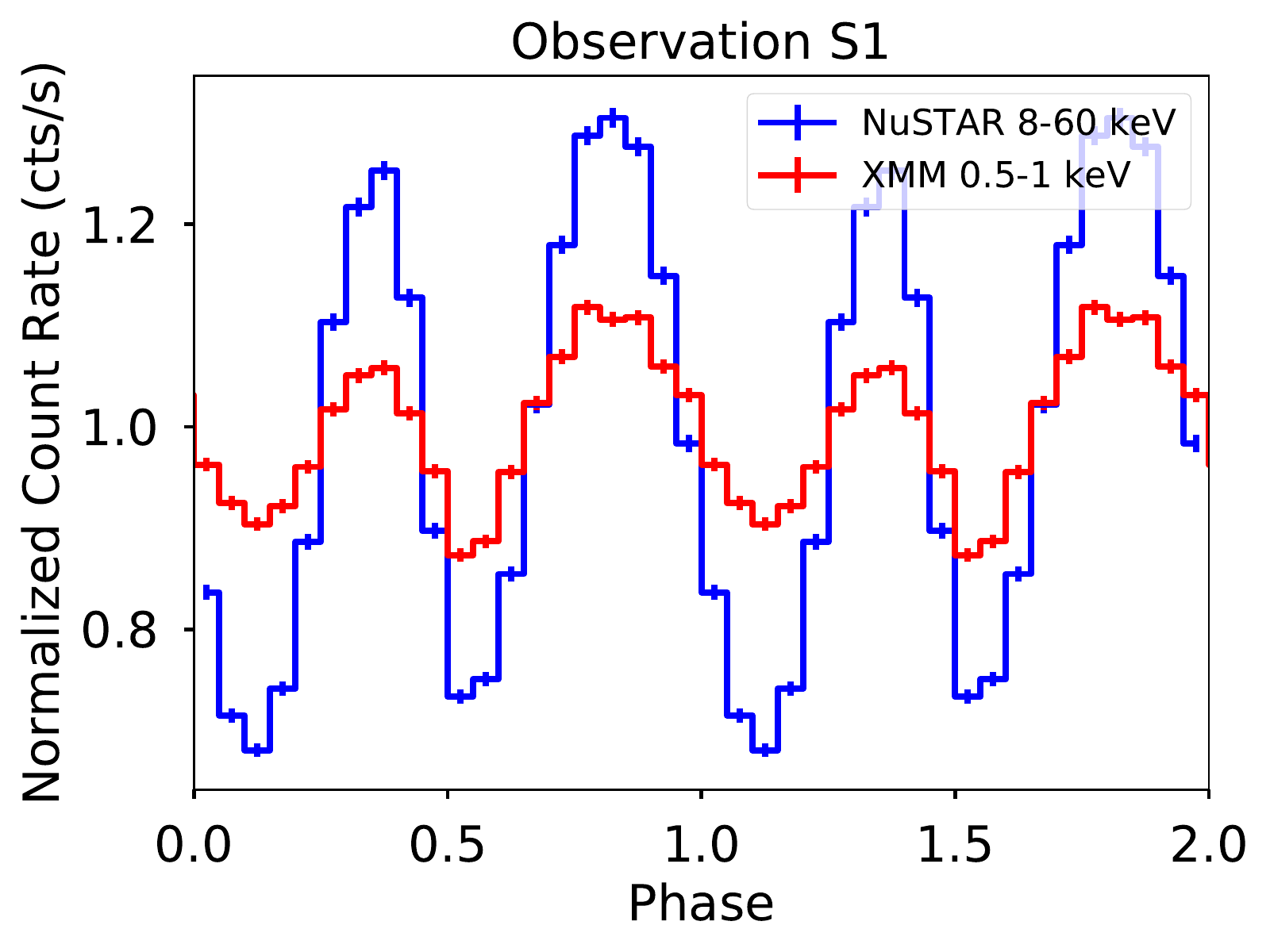} & \includegraphics[scale=0.45]{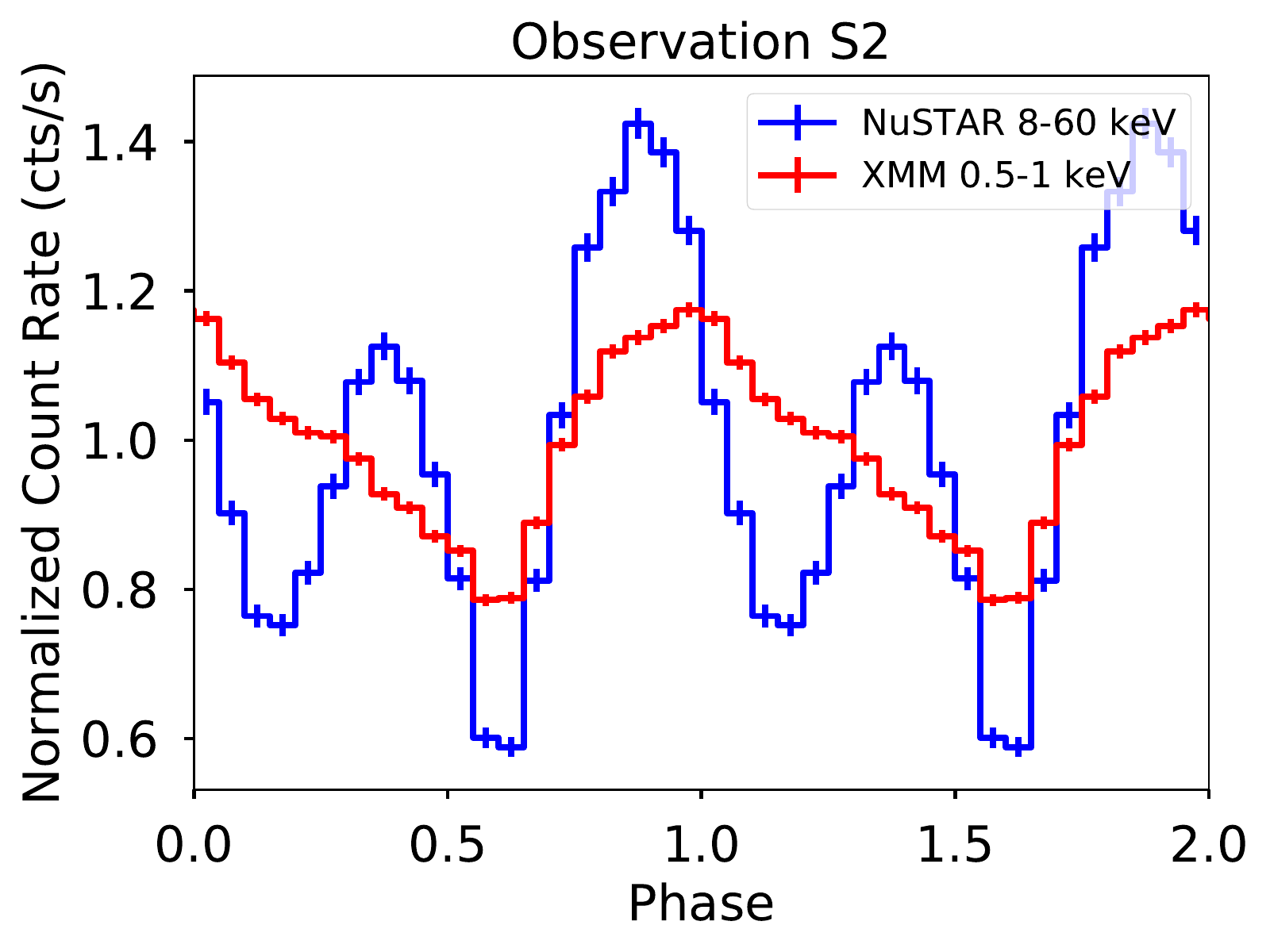} \\
	\multicolumn{2}{c}{\includegraphics[scale=0.45]{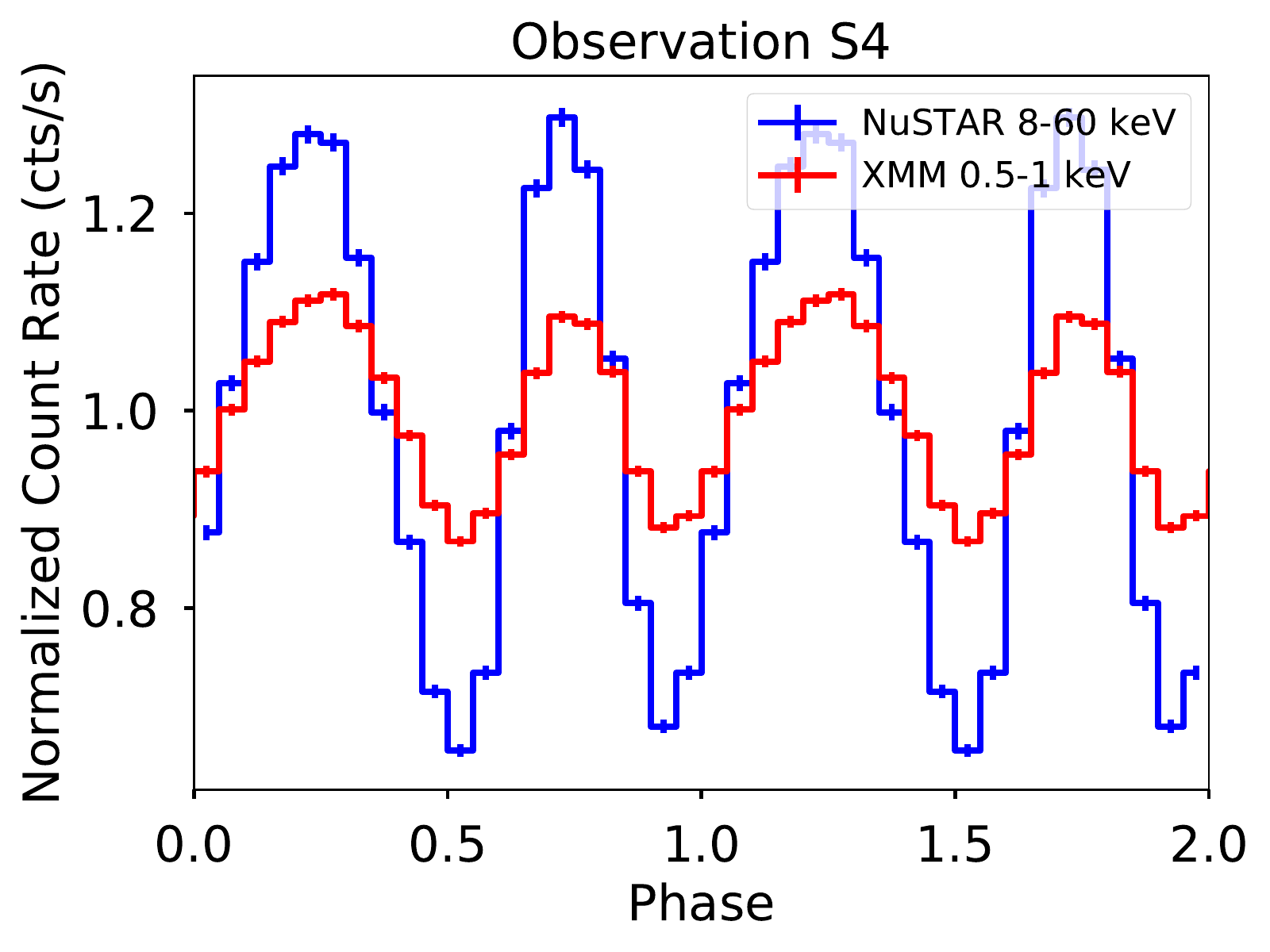} }
\end{tabular}
\caption{Same as Figure \ref{fig:lmcpp} for the three SMC X-1 observations with detected pulsations. The pulse profiles for Observations S1 and S4 are almost identical, which is consistent with covering a complete precession cycle of the inner accretion disk.}
\label{fig:smcpp}
\end{figure*}

We filtered the SMC X-1 data by energy in the same way as the LMC X-4 data so that our \nustar\ pulse profile captures the hard X-ray component and our \xmms\ data covers the soft component. We then made pulse profiles with 20 bins per phase (Figure \ref{fig:smcpp}) using the Stingray (\citealt{stingray}) software tool {\fontfamily{qcr}\selectfont fold\_events} and the measured period and period derivative.

\begin{deluxetable} {ccc} 
\tablecolumns{3}
\tablecaption{Best Fit Spin Periods for SMC X-1 Observations}  
\tablewidth{0pt}
\tablehead{
\colhead{Observation} & \colhead{Spin Period (ms)} & \colhead{$\dot{P}$ (ss$^{-1}$) } }
\startdata
S1 & 699.65 $\pm$ 0.03 & (-1 $\pm$ 3)$\times10^{-9}$ \\
S2 & 699.59 $\pm$ 0.04 &(1 $\pm$ 3)$\times10^{-10}$ \\
S4 & 699.60 $\pm$ 0.03 & (3 $\pm$ 3)$\times10^{-10}$
\enddata
\label{tab:smcperiod}
\end{deluxetable}

\subsection{Spectral Analysis}
\subsubsection{Phase-averaged Spectroscopy}

For both data sets, we extracted spectra from the source and background regions described above using appropriate NuSTARDAS and XMMSAS selection tools. However, we did not select background spectra for the \xmm\ observations of SMC X-1 because the source flux dominates the EPIC-pn Timing Mode CCD (e.g.\ \citealt{ng2010}).

We grouped all \nustar\ spectra into bins with a signal to noise ratio of 18 and all \xmm\ spectra with a minimum of 100 counts per bin, which produced good statistics. We fitted the phase-average spectra in the range of 0.6--50 keV.

We modeled the spectra in Xspec version 12.9.1 (\citealt{arnaud1996}). If possible, we wished to apply the same continuum model to both LMC X-4 and SMC X-1 spectra to allow for a direct comparison. We tested continuum models including Negative and Positive EXponential (NPEX, e.g.\ \citealt{mihara1998}), a power law with a Fermi-Dirac cutoff (FDCut, \citealt{tanaka1986}), and a power law with a high energy cutoff (\citealt{white1983}). We found that the FDCut and high energy cutoff had slightly higher reduced $\chi^{2}$ values and large residuals at high energies. For these reasons, we selected NPEX as our best continuum model. In Xspec, our NPEX model was defined as  
$$ f(E) = n_{1}(E^{-\alpha_{1}} + n_{2}E^{-\alpha_{2}}) e^{-E/kT},$$
where we fixed $\alpha_{2} = -2$.

In addition to NPEX, our spectral model also included an absorbing column (tbnew), a blackbody with $kT \sim 0.17$ keV, and several Gaussian emission lines at 6.4 keV (Fe K$\alpha$), 1.02 keV (Ne X Ly$\alpha$), 0.91 keV (Ne IX), and 0.65 keV (O VIII Ly$\alpha$). Each of these emission lines has been previously detected in LMC X-4 spectra with the \textit{Chandra} High Energy Transmission Grating Spectrometer (\citealt{neilsen2009}) and in SMC X-1 spectra with the \textit{Chandra} ACIS instrument (\citealt{vrtilek2001,vrtilek2005}). To reduce degeneracy in the blackbody model components, we fixed the widths of the Ne X Ly$\alpha$, Ne IX, and O VIII Ly$\alpha$ lines to the values found by \cite{neilsen2009}. In all observations, we found the Fe K$\alpha$ line was quite broad and that a 0.5 keV line width provided a good fit. However, in Observation S4 the spectrum also required a narrow (0.1 keV) component, as also seen by \cite{neilsen2009}.

To reduce degeneracies between the absorption and the blackbody component, we fixed the absorbing column density to the Galactic value in the direction of our sources, which we calculated using the HI4PI Map (\citealt{hi4pi2016}) via the HEASARC \nh\ calculator. These values were 1$\times 10^{21}$ cm$^{-2}$ for LMC X-4 and 3$\times 10^{21}$ cm$^{-2}$ for SMC X-1.

For the LMC X-4 spectra, we used the elemental abundances described in \cite{hanke2010} to account for the LMC's lower metallicity relative to Galactic abundances. The SMC X-1 spectral fits were performed using abundances from \cite{wilms2000}. For both sources, we used the cross sections from \cite{verner1996}.

\begin{figure*}
\centering
\begin{tabular}{c}
	\includegraphics[scale=0.73]{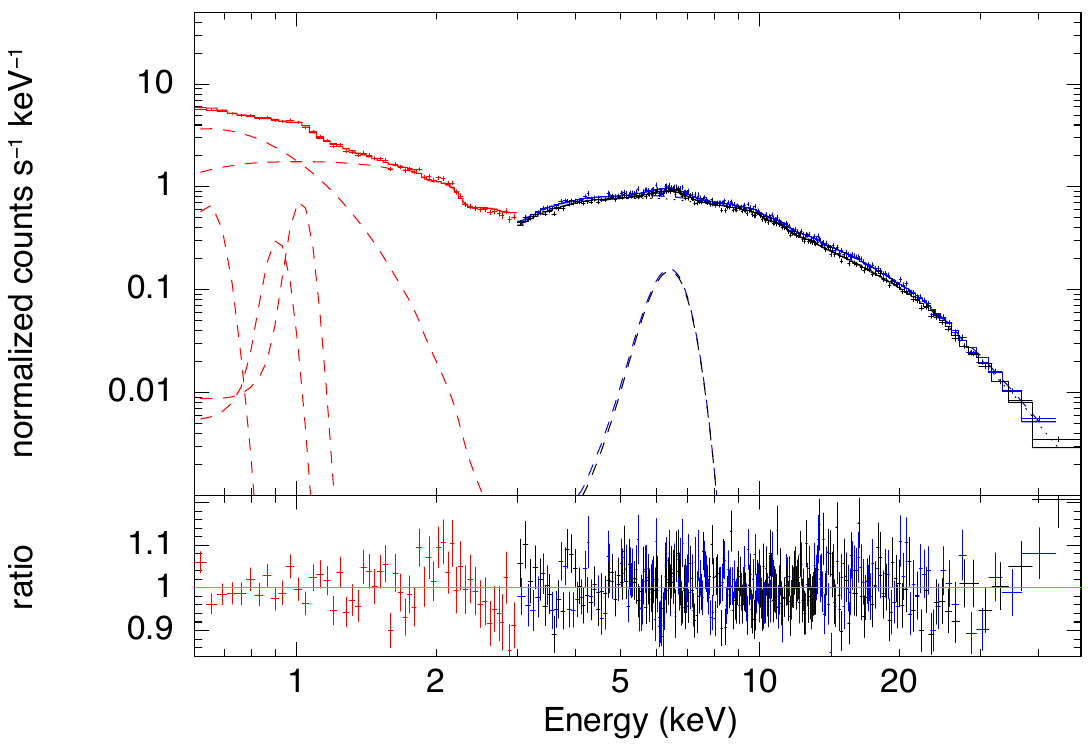} \\
	 \includegraphics[scale=0.3]{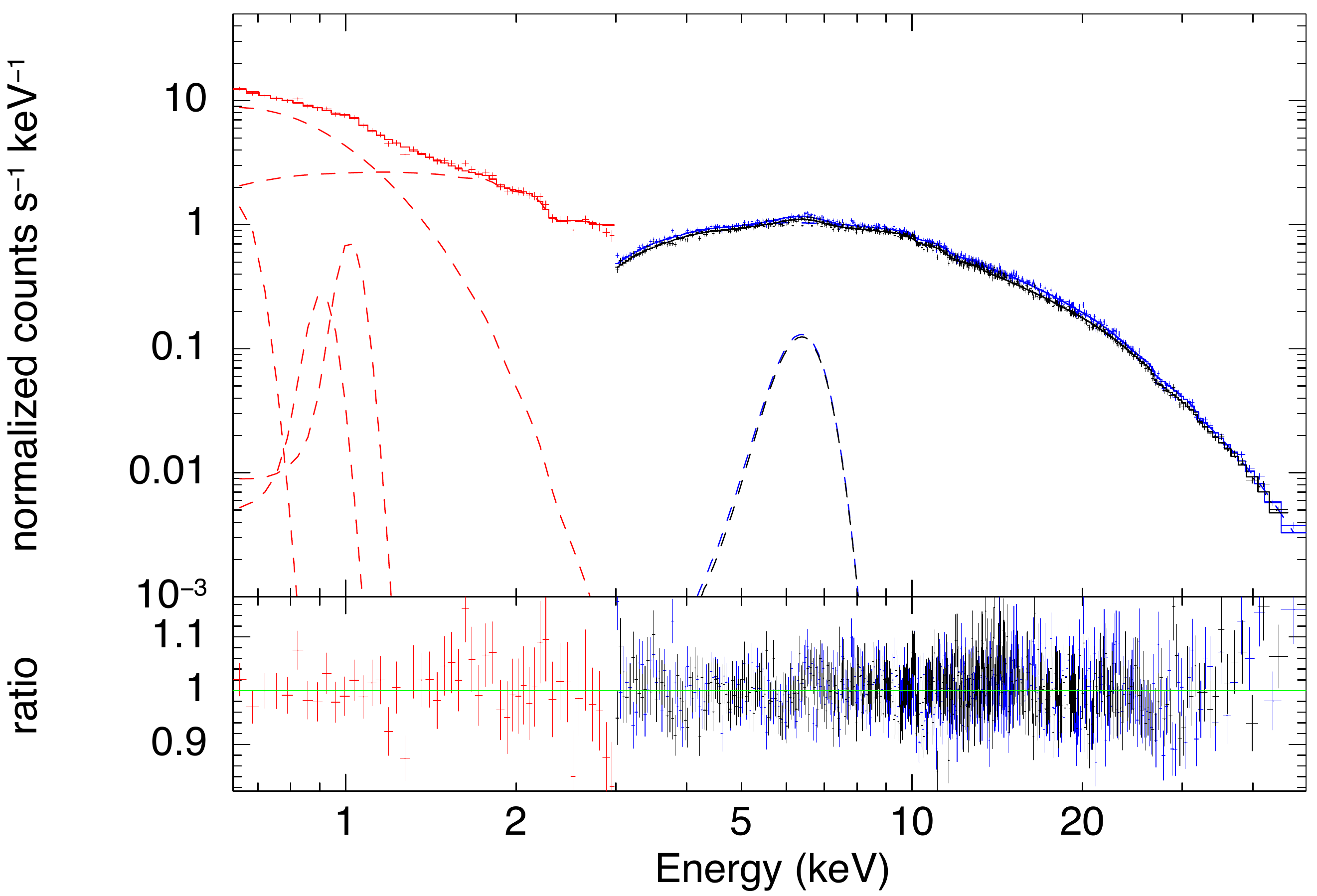} \\
	 \includegraphics[scale=0.3]{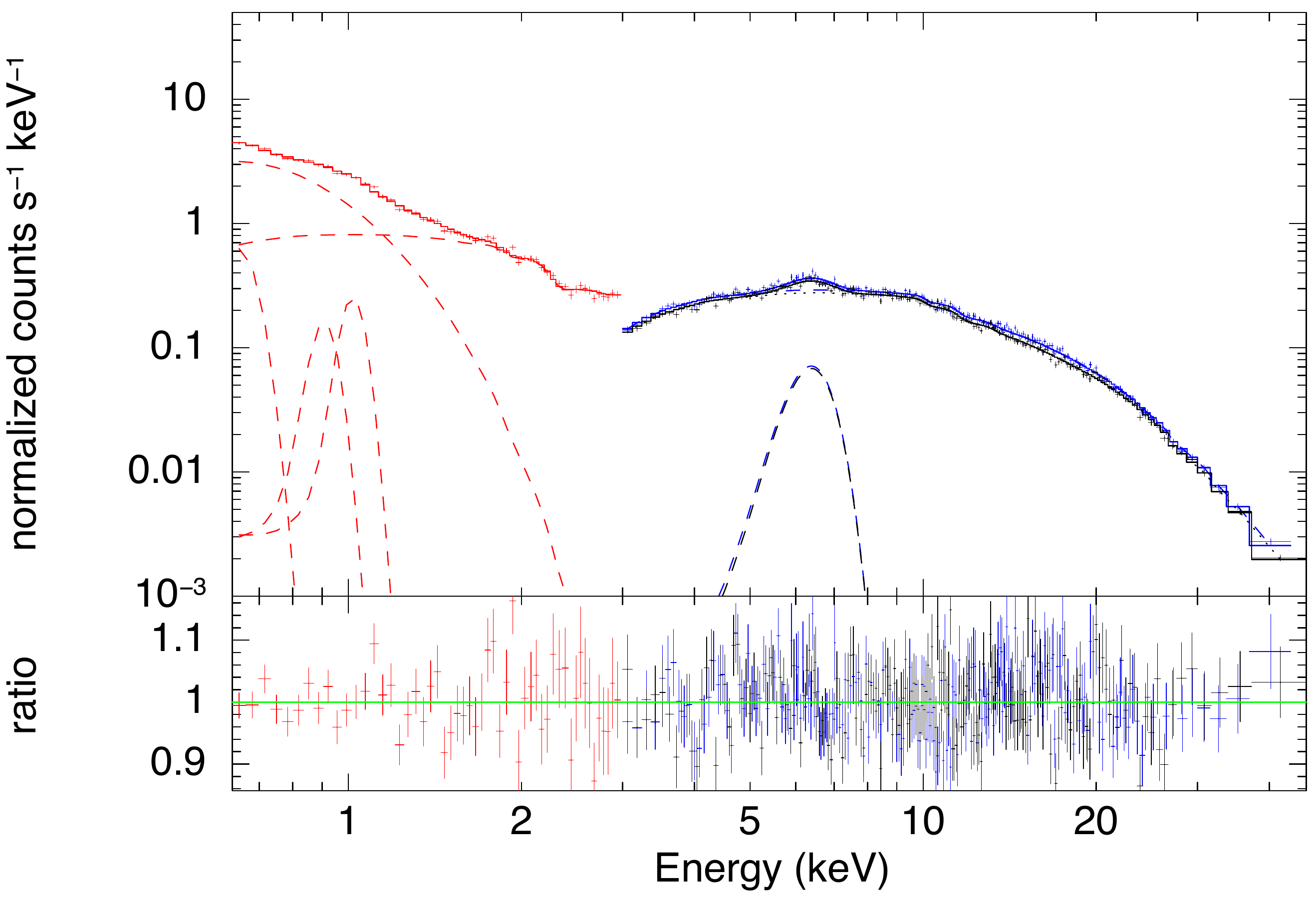} \\
	\includegraphics[scale=0.3]{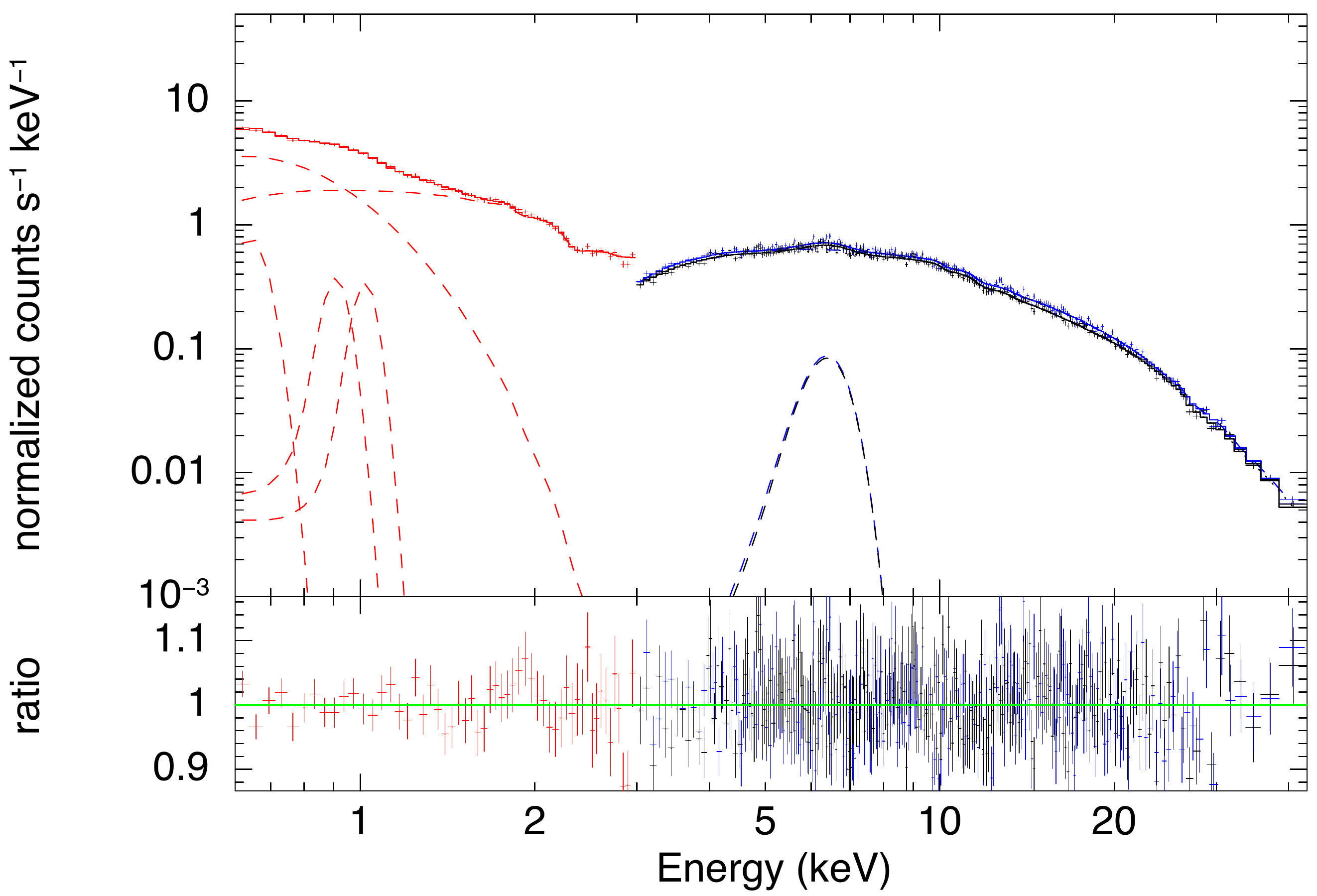} 
\end{tabular}
\caption{Joint \xmm\ (red) and \nustar\ (FPMA - blue, FPMB - black) spectra for the four LMC X-4 observations. The \xmm\ spectrum is modeled from 0.6--3 keV while the \nustar\ spectra are modeled from 3--50 keV. The ratios of data to model are plotted below each spectra. Spectral parameters for all four observations are located in Table \ref{tab:lnpexparams}. }
\label{fig:lspec}
\end{figure*}

\begin{figure*}
\centering
\begin{tabular}{c}
	\includegraphics[scale=0.73]{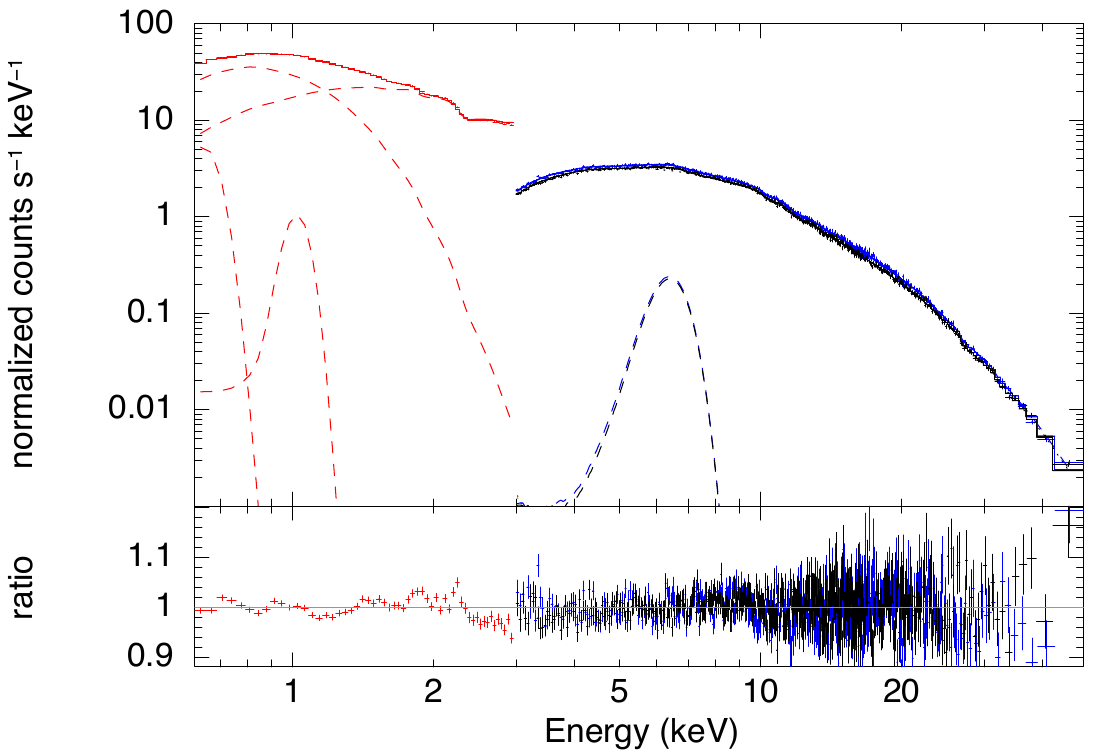} \\
	 \includegraphics[scale=0.3]{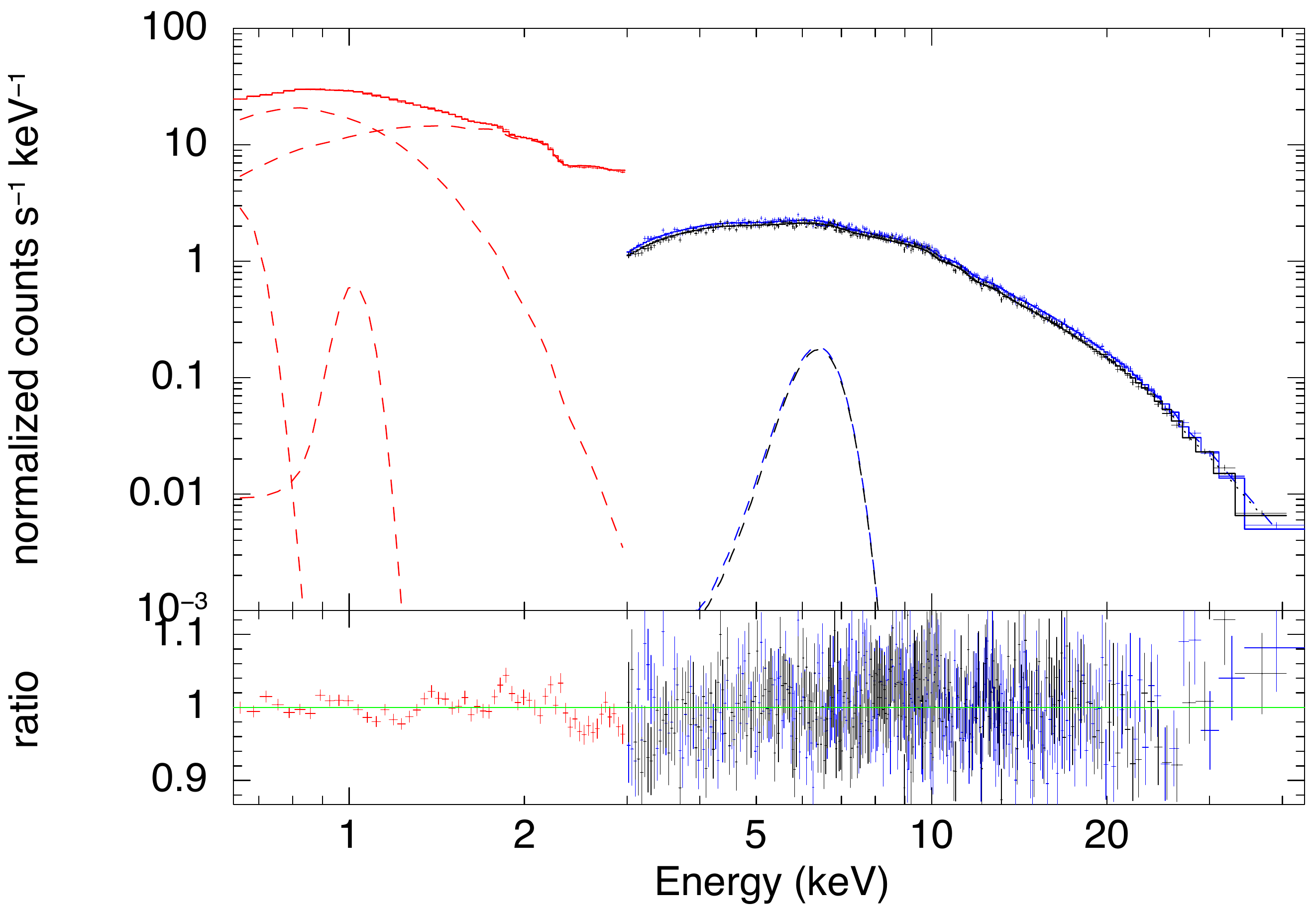} \\
	\includegraphics[scale=0.3]{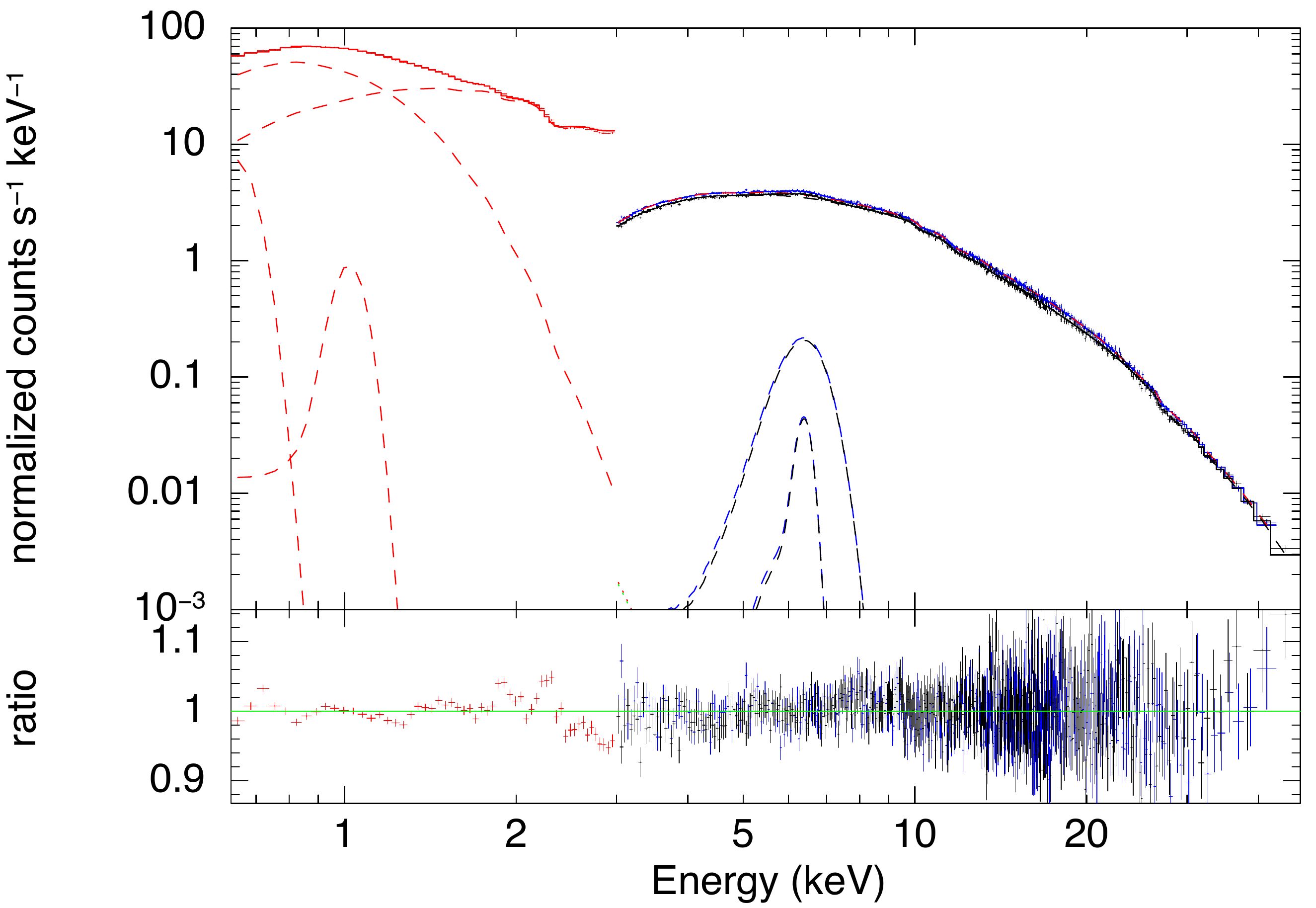} 
\end{tabular}
\caption{Joint \xmm\ (red) and \nustar\ (FPMA - blue, FPMB - black) spectra for the three SMC X-1 observation. The \xmm\ spectrum is modeled from 0.6--3 keV while the \nustar\ spectra are modeled from 3--50 keV. These spectra did not require the Ne IX emission line at 0.91 keV, and so we removed it from the model. Observation S4 contained both a broad and narrow Fe K$\alpha$ line. The spectral parameters for these observations can be found in Table \ref{tab:snpexparams}.}
\label{fig:sspec}
\end{figure*}

The phase averaged spectra and the residuals to the model fit for both data sets are shown in Figures \ref{fig:lspec} and \ref{fig:sspec}. The spectral parameters and their uncertainties are given in Tables \ref{tab:lnpexparams} and \ref{tab:snpexparams}. We chose not to model \xmm\ and \nustar\ in overlapping energy ranges to improve the model fit by minimizing differences in the response functions from these two observatories.

For all LMC X-4 and SMC X-1 spectra, we fit the models jointly to the \xmms, \nustar\ FPMA and FPMB spectra. The parameters in the \nustar\ spectra are tied to those in the \xmms\ spectrum via a cross-calibration constant that accounts for differences in observed flux between the telescopes. The constants for \nustar\ FPMA and FPMB are in good agreement in all spectra. However, as can be seen in Tables \ref{tab:lnpexparams} and \ref{tab:snpexparams}, there is a discrepancy between the \xmms\ and \nustar\ calibration constants, with the \nustar\ constants being 2--3 the \xmms\ constant (when $c_{XMM}=1$). This issue was even more pronounced for the SMC X-1 spectra, where \xmm\ was in Timing Mode. We investigated the cross normalization in detail, and found our choice of wide \xmm\ source extraction regions and non-overlapping energy when fitting drove the cross normalization unrealistically high. We verified that changes in source extraction region did not impact the spectral or pulse profile shapes. We modeled the joint \nustar\ and \xmm\ spectra in the overlapping 3--10 keV range and found that the \nustar\ and \xmm\ flux in this energy range agreed within 10\%. A full exploration of the cross normalization is outside the scope of this work, however we are confident that the values shown in Tables \ref{tab:lnpexparams} and \ref{tab:snpexparams} are a reflection of our analysis steps and do not reflect the relative fluxes measured by \xmm\ and \nustar.

\begin{deluxetable*}{lcccc}
\tablecolumns{5}
\tablecaption{LMC X-4 phase-averaged spectral parameters\tablenotemark{a}}
\tablewidth{0pt}
\tablehead{
\colhead{Parameter} & \colhead{Observation L1} & \colhead{Observation L2} & \colhead{Observation L3} & \colhead{Observation L4} }  
\startdata
$kT_{\text{BB}}$ (keV) 			&  0.168 $\pm$ 0.006			&0.168 $\pm$ 0.006				& 0.161 $\pm$ 0.004 		& 0.160 $\pm$ 0.006\\
$A_{\text{BB}}$ (keV) 			&  (5.2 $\pm$ 0.3)$\times 10^{-4}$	&(7.0 $\pm$ 0.4)$\times 10^{-4}$	& (2.27 $\pm$ 0.09)$\times 10^{-4}$ & (3.2 $\pm$ 0.2)$\times 10^{-4}$\\
$\alpha_{1}$ 					&  0.55 $\pm$ 0.03				&0.41 $\pm$ 0.02				& 0.55 $\pm$ 0.06			& 0.66 $\pm$ 0.04\\
$A_{\alpha_{2}}$ 				&  (2.3 $\pm$ 0.1)$\times 10^{-3}$ 	& (4.2 $\pm$ 0.1)$\times 10^{-3}$	& (5.5 $\pm$ 0.3)$\times 10^{-3}$  & (2.9 $\pm$ 0.1) $\times 10^{-3}$\\ 
$kT_{\text{fold}}$ (keV) 			&  6.1 $\pm$ 0.1				&6.26 $\pm$ 0.05				&  5.81 $\pm$ 0.07  			& 6.21 $\pm$ 0.07\\
log$_{10}$($F_{3-40 \text{ keV}}$) 	&  -9.49 $\pm$ 0.01	 			&-9.38 $\pm$ 0.01				& -9.97 $\pm$ 0.02			& -9.66 $\pm$ 0.01\\
E$_{\text{Fe K$\alpha$}}$ (keV, fixed) 	& 6.4						&6.4 							& 6.4 					& 6.4\\
$\sigma_{\text{Fe K$\alpha$}}$ (keV, fixed) & 0.5					&0.5 							& 0.5   					& 0.5\\
$A_{\text{Fe K$\alpha$}}$  		& (3.3 $\pm$ 0.3)$\times 10^{-4}$	&(2.2 $\pm$ 0.2)$\times 10^{-4}$	& (1.04 $\pm$ 0.09)$\times 10^{-4}$  & (1.3 $\pm$ 0.2)$\times 10^{-4}$\\
$E_{\text{Ne X Ly$\alpha$}}$ (keV, fixed) 	&  1.02					& 1.02						& 1.02					& 1.02 \\
$\sigma_{\text{Ne X Ly$\alpha$}}$ (keV, fixed) 	& 0.003				& 0.003						& 0.003					& 0.003 \\
$A_{\text{Ne X Ly$\alpha$}}$ 		&(7 $\pm$ 1)$\times 10^{-4}$		&(4 $\pm$ 2)$\times 10^{-4}$		&(1.2 $\pm$ 0.4)$\times 10^{-4}$	&(1.9 $\pm$ 0.7)$\times 10^{-4}$ \\
$E_{\text{Ne IX}}$ (keV, fixed) 		&  0.91						&0.91						& 0.91  					& 0.91\\
$\sigma_{\text{Ne IX}}$ (keV, fixed) 	& 0.003						&0.003						&  0.22 $\pm$ 0.01 			& 0.20 $\pm$ 0.01\\
$A_{\text{Ne IX}}$  				& (3 $\pm$ 2)$\times 10^{-4}$		& (1 $\pm$ 2)$\times 10^{-4}$		& (8 $\pm$ 5)$\times 10^{-5}$	& (2.2 $\pm$ 0.8)$\times 10^{-4}$\\
$E_{\text{O VIII Ly$\alpha$}}$ (keV, fixed) 	& 0.65				& 0.65						& 0.65					& 0.65 \\
$\sigma_{\text{O VIII Ly$\alpha$}}$ (keV, fixed) & 0.003				& 0.003						& 0.003					& 0.003 \\
$A_{\text{O VIII Ly$\alpha$}}$ 		&(1.0 $\pm$ 0.4)$\times 10^{-4}$	&(1.1 $\pm$ 0.6)$\times 10^{-3}$	& (5 $\pm$ 1)$\times 10^{-4}$	&(7 $\pm$ 2)$\times 10^{-4}$ \\
$c_{\text{EPIC-pn}}$ (fixed)		&  1							&1							&	1	 				& 1\\
$c_{\text{FPMA}}$	 			& 2.10 $\pm$ 0.07				&2.58 $\pm$ 0.08				& 3.0 $\pm$ 0.1			& 2.99 $\pm$ 0.09\\
$c_{\text{FPMB}}$ 				& 2.13 $\pm$ 0.07				&2.65 $\pm$ 0.09				&  3.1 $\pm$ 0.1 			& 3.09 $\pm$ 0.09 \\
\chisq\ 						& 483.05						&941.03						&  478.50 					& 612.03\\
Degrees of Freedom 			& 440						&908							&  425					& 568
\enddata
\tablenotetext{a}{For the continuum model {\fontfamily{qcr}\selectfont constant * tbnew * (cflux * npex + bbody + gauss + gauss + gauss + gauss)}. All errors are 90\% confidence intervals.}
\label{tab:lnpexparams}
\end{deluxetable*}

\begin{deluxetable*}{lccc}
\tablecolumns{4}
\tablecaption{SMC X-1 phase-averaged spectral parameters\tablenotemark{a}}
\tablewidth{0pt}
\tablehead{
\colhead{Parameter} & \colhead{Observation S1} & \colhead{Observation S2} & \colhead{Observation S4} }  
\startdata
$kT_{\text{BB}}$ (keV) 				&  0.182 $\pm$ 0.001				&0.179 $\pm$ 0.002				& 0.184 $\pm$ 0.001  \\
$A_{\text{BB}}$ (keV) 				&  (1.45 $\pm$ 0.01)$\times 10^{-3}$		&(8.50 $\pm$ 0.09)$\times 10^{-4}$	& (2.06 $\pm$ 0.01)$\times 10^{-3}$ \\
$\alpha_{1}$ 						&  0.402 $\pm$ 0.007				&0.44 $\pm$ 0.02				& 0.397 $\pm$ 0.007  \\
$A_{\alpha_{2}}$ 					&  (1.01 $\pm$ 0.07)$\times 10^{-3}$ 	&(1.9 $\pm$ 0.2)$\times 10^{-3}$ 	& (1.18 $\pm$ 0.08)$\times 10^{-3}$  \\ 
$kT_{\text{fold}}$ (keV) 				&  5.64 $\pm$ 0.07					&5.2 $\pm$ 0.1  				&  5.56 $\pm$ 0.07  \\
log$_{10}$($F_{3-40 \text{ keV}}$) 		&  -9.260 $\pm$ 0.003 				&-9.450 $\pm$ 0.006				& -9.119 $\pm$ 0.003 \\
E$_{\text{Fe K$\alpha$, broad}}$ (keV, fixed) 	& 6.4								&6.4 							& 6.4 \\
$\sigma_{\text{Fe K$\alpha$, broad}}$ (keV, fixed) & 0.5							&0.5 							& 0.5   \\
$A_{\text{Fe K$\alpha$, broad}}$  			& (3.3$\pm$ 0.5)$\times 10^{-4}$		&(2.4 $\pm$ 0.3)$\times 10^{-4}$	& (3.6 $\pm$ 0.6)$\times 10^{-4}$  \\
E$_{\text{Fe K$\alpha$, narrow}}$ (keV, fixed) 	& N/A							&N/A 						& 6.4 \\
$\sigma_{\text{Fe K$\alpha$, narrow}}$ (keV, fixed) & N/A							& N/A						& 0.1 (fixed)  \\
$A_{\text{Fe K$\alpha$, narrow}}$  			& N/A							&N/A							& (2 $\pm$ 3)$\times 10^{-5}$  \\
$E_{\text{Ne X Ly$\alpha$}}$ (keV, fixed) 		&  1.02							&1.02						& 1.02   \\
$\sigma_{\text{Ne X Ly$\alpha$}}$ (keV, fixed) 	& 0.003							&0.003						& 0.003\\
$A_{\text{Ne X Ly$\alpha$}}$  				& (2.2 $\pm$ 0.7)$\times 10^{-4}$		&(1.3 $\pm$ 0.5)$\times 10^{-4}$	& (2.0 $\pm$ 0.8)$\times 10^{-4}$ \\
$E_{\text{O VIII Ly$\alpha$}}$ (keV, fixed) 	& 0.65							&0.65						&0.65 \\
$\sigma_{\text{O VIII Ly$\alpha$}}$ (keV, fixed) & 0.003							&0.003						&0.003 \\
$A_{\text{O VIII Ly$\alpha$}}$ 				& (3.1$\pm$ 0.2)$\times 10^{-3}$		&(1.7 $\pm$ 0.5)$\times 10^{-3}$	& (4.3 $\pm$ 0.3)$\times 10^{-3}$ \\
$c_{\text{EPIC-pn}}$ (fixed)			&  1								&1							&	1	 \\
$c_{\text{FPMA}}$	 				& 3.21 $\pm$ 0.02					&3.40 $\pm$ 0.05				& 2.63 $\pm$ 0.02	 	 \\
$c_{\text{FPMB}}$ 					& 3.26 $\pm$ 0.03					&3.48 $\pm$ 0.05				&  2.68 $\pm$ 0.02 \\
\chisq\ 							& 1317.43							&725.34						&  1452.61  \\
Degrees of Freedom 				& 963							&614							& 973	
\enddata
\tablenotetext{a}{For the continuum model {\fontfamily{qcr}\selectfont constant * tbnew * (cflux * npex + bbody + gauss + gauss + gauss + gauss)}. All errors are 90\% confidence intervals.}
\label{tab:snpexparams}
\end{deluxetable*}

\begin{figure*}
\centering
\begin{tabular}{c}
	\includegraphics[scale=0.73]{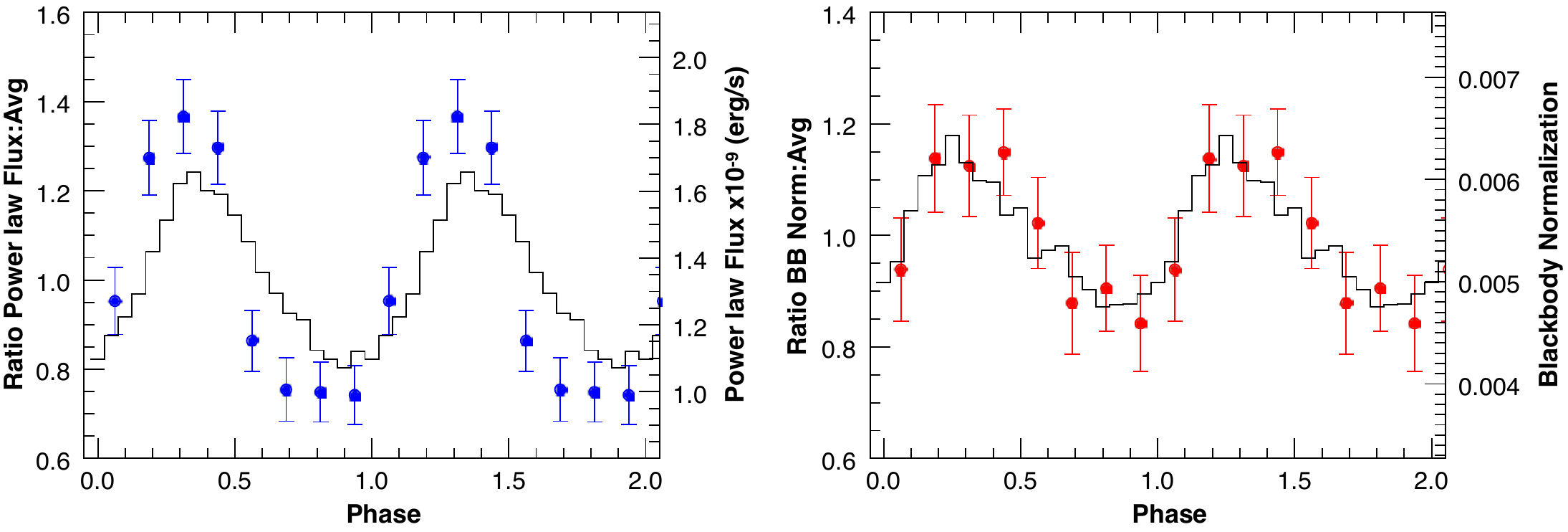} \\
\end{tabular}
\caption{Left: Power law flux (3--40 keV, blue points) plotted together with the \nustar\ 8--60 pulse profile for Observation L3. Right: Blackbody normalization (red points) plotted against the \xmm\ 0.5--1 pulse profile for Observation L3. In both figures, the spectral parameters are in good agreement with their respective pulse profiles. We find similarly good agreement in all LMC X-4 and SMC X-1 observations, but merely show this observation as an example. The agreement of the blackbody and power law spectral parameters with their respective energy resolved pulse profiles indicates that the pulse profiles are a suitable proxy for these spectral parameters in our warped disk model.}
\label{fig:ppsfluxbb}
\end{figure*}

\subsubsection{Phase-resolved Spectroscopy}

We also performed phase-resolved spectroscopy for all LMC X-4 and SMC X-1 observations. The phase-averaged spectra were filtered into 8 equal phase bins by using the HENDRICS tool {\fontfamily{qcr}\selectfont HENphasetag} to calculate spin phase for each photon. The \nustar\ spectra were filtered using {\fontfamily{qcr}\selectfont xselect} and the \xmm\ spectra were filtered using XMMSAS. All spectra were grouped to have a minimum of 100 counts per spectral bin. We fitted the phase-resolved spectra in the range of 0.6--40 keV.

We used the same model as for the phase-average spectra when fitting the phase-resolved spectra; however, to reduce the number of free parameters we fixed the blackbody temperatures to their respective phase-averaged values. We also found that the O VIII, Ne IX, and Ne X emission lines were not required and poorly constrained by the lower resolution phase-resolved spectra. We removed these lines from the phase-resolved model spectra.

Other than these changes to the spectral model, the phase-resolved spectra were fit using the same methods, abundances, and energy ranges specified for the phase-averaged spectroscopy.

\section{Results} \label{sec:results}

\subsection{Pulse Profiles} 

The LMC X-4 and SMC X-1 pulse profiles shown in Figures \ref{fig:lmcpp} and \ref{fig:smcpp} show changes in shape and phase over the course of a single superorbital cycle.

For LMC X-4, the hard (8--60 keV) and soft (0.5--1 keV) pulse profiles from Observations L1 and L2 are out of phase. In Observation L1 the soft pulses are slightly less than 180\degree\ out of phase, while in Observation L2 they appear to be closer to 180\degree\ out of phase. By contrast, the hard and soft pulse profiles in Observation L3 are almost completely in phase. The pulse profiles, and in particular the hard pulses, in Observation L4 are weakly detected due to the pulse dropout phenomenon that occurred during this observation (see \citealt{brumback2018b}). Despite this, we observe that the hard and soft pulsations appear out of phase.

Independent of pulse dropout behaviors, we also observe changes in pulse shape with superorbital phase in LMC X-4. The hard pulse profiles in Observations L1 and L3 are relatively smooth single peaks, while the hard pulse profile in Observation L2 has become broad and flat. In general, the soft pulse shapes in all LMC X-4 observations are rounded single peaks. We also observe a shift in relative strength between hard and soft pulsations in Observation L2, apparently driven by a change in the hard pulsed fraction.

The pulse profile for SMC X-1 is double peaked. With the energy-resolved pulse profiles for SMC X-1, we find that the profiles for Observations S1 and S4 are extremely consistent in shape and relative phase; both hard and soft profiles are in phase with each other and both show two peaks of approximately equal strength in both the hard and soft pulses. These shapes are different than those seen in Observation S3, where the hard pulses show one strong and one weak peak, while the soft pulses have merged into a broad single peak.

Because the hard pulsations are caused by the pulsar beam and the soft pulsations originate from accretion disk reprocessing (e.g.\ \citealt{hickox2004}), the consistency in pulse phase between observations from the same superorbital phase, particularly Observations S1 and S4, shows that we have observed a complete precession cycle of the inner accretion disk.

\subsection{Spectroscopy}
Our spectroscopic analysis of these data indicates that the broad-band X-ray spectra of LMC X-4 and SMC X-1 are well described by an absorbed power law and a soft blackbody component. In both sources, the blackbody temperature changes very little with superorbital phase. In SMC X-1, we also find very little variation in the $\alpha_{1}$ parameter with superorbital phase. There is, however, some variation in the strength of the second power law ($A_{\alpha_{2}}$) which indicates that the overall shape of the hard continuum is changing slightly with superorbital phase. In LMC X-4, we observe changes in power law shape through variation in both $\alpha_{1}$ and $A_{\alpha_{2}}$. 

We would expect that Observations S1 and S4 would have generally the same spectral shape since these observations were taken at the same superorbital phase, and the same applies to Observations L1 and L4. We do find good agreement between the spectral parameters in Observations S1 and S4, where the only notable differences are a slightly stronger second power law normalization and the presence of a narrow Fe K$\alpha$ feature in Observation S4. We find less good agreement between the spectral parameters in Observations L1 and L4; however, we do not consider these discrepancies to be problematic considering that these two observations sample different accretion and pulse behaviors (\citealt{brumback2018b}). In Observation L1, our phase-averaged spectrum reflects a strongly pulsed time interval between bright accretion flares, whereas in Observation L4 our spectrum is drawn from a weakly pulsed pre-flare interval. The hardness ratios vary between these two states, implying that the shape of the spectrum changes (see Fig. 1 in \citealt{brumback2018b}). In \cite{brumback2018b} we suggest that these different pulse behaviors could be driven by changing emission geometries during the accretion flares. If this is indeed the case, we would expect to see differences in the spectral shape during this process.

\begin{deluxetable*} {ccccc} 
\tablecolumns{5}
\tablecaption{Disk Model Parameters}  
\tablewidth{0pt}
\tablehead{ \multicolumn{1}{c}{}  & \multicolumn{2}{c}{LMC X-4}  & \multicolumn{2}{c}{SMC X-1} \\ 
\cmidrule(lr){2-3} \cmidrule(lr){4-5} \\ 
\colhead{Parameter} & \colhead{Pencil Beam} & \colhead{Fan Beam} & \colhead{Pencil Beam} & \colhead{Fan Beam} }
\startdata
$r_{\text{in}}$ (10$^{8}$ cm) & 0.8 & 0.8 & 0.8 & 0.8 \\
$r_{\text{out}}$ (10$^{8}$ cm) & 1 & 1 & 1 & 1 \\
Inner tilt $\theta_{\text{in}}$ (\degree) & 10 & 10 & 10 & 10\\
Outer tilt $\theta_{\text{out}}$ (\degree) & 45 & 45 &  45 & 45\\
Twist angle $\phi_{\text{tw}}$ (\degree) & -130 & -130 & -130 & -130\\
Beam$_{1}$ angle from rotational plane $\theta_{\text{b}1}$ (\degree) 	& 	60\tablenotemark{b,c,d}, 75\tablenotemark{a}		&  60\tablenotemark{c}, 70\tablenotemark{b,d}, 75\tablenotemark{a} 		& 60\tablenotemark{e,g}, -50\tablenotemark{f}		& 60\tablenotemark{e,g}, -40\tablenotemark{f}\\
Beam$_{2}$ angle from rotational plane $\theta_{\text{b}2}$ (\degree) 	& 60\tablenotemark{b,d}, 65\tablenotemark{a}, -60\tablenotemark{c} 		& 60\tablenotemark{a}, 70\tablenotemark{b,d}, -60\tablenotemark{c}, 	& 60\tablenotemark{e,f,g} 					& 60\tablenotemark{e,f,g}\\
Beam$_{1}$ azimuth $\phi_{\text{b}1}$ (\degree) & 0 & 0 & 0 & 0 \\
Beam$_{2}$ azimuth $\phi_{\text{b}2}$ (\degree) & 110\tablenotemark{a}, 130\tablenotemark{b,d}, 160\tablenotemark{c}	 & 110\tablenotemark{a},120\tablenotemark{b,d}, 160\tablenotemark{c} 		&180\tablenotemark{e,g}, 185\tablenotemark{f}	& 180\tablenotemark{e,f,g} \\
Beam half-width $\sigma_{\text{b}}$ (\degree) & 30\tablenotemark{a}, 45\tablenotemark{b,d}, 60\tablenotemark{c} & 30 & 60 & 30 \\
Fan beam opening angle $\theta_{\text{fan}}$ (\degree) & 0 & 15\tablenotemark{a}, 20\tablenotemark{c}, 25\tablenotemark{b,d} & 0 & 30\tablenotemark{e,f,g}\\
Observer elevation $\theta_{\text{obs}}$ (\degree) & 40 & 40 & 20 & 20
\enddata
\tablenotetext{a}{This value required for Observation L1}
\tablenotetext{b}{This value required for Observation L2}
\tablenotetext{c}{This value required for Observation L3}
\tablenotetext{d}{This value required for Observation L4}
\tablenotetext{e}{This value required for Observation S1}
\tablenotetext{f}{This value required for Observation S2}
\tablenotetext{g}{This value required for Observation S4}
\label{tab:diskpar}
\end{deluxetable*}

In our phase-resolved analysis, the parameters allowed to vary within each spectrum were the blackbody normalization, the overall flux of the power law, the primary power law index $\alpha_{1}$, the secondary power law normalization, the power law folding energy, and the Fe K$\alpha$ line normalization. Across the phase-resolved spectra for both LMC X-4 and SMC X-1, we only find clear, coherent changes with pulse phase in the blackbody normalization, the power law flux, and the Fe K$\alpha$ line normalization. The other parameters ($\alpha_{1}$, folding temperature, and second power law normalization) are either consistent with being constant, or show variations that are difficult to describe physically due to degeneracies within the model and reduced signal to noise in the spectra. 

An example of the smooth variations in power law flux and blackbody normalization are shown in Figure \ref{fig:ppsfluxbb} for Observation L3, where the variation in power law flux is overplotted with the hard \nustar\ pulse profile, and the blackbody normalization is overplotted with the soft \xmm\ pulse profile. The other LMC X-4 and SMC X-1 observations show similarly good agreement between these two spectral parameters and the pulse profiles, and so for the sake of brevity we do not show them here. The agreement of these parameters with their respective pulse profiles is significant because it indicates that the pulse profiles (measured in count rates) are reasonable proxy for the spin-resolved power law and blackbody flux. This agreement allows us to directly fit the energy resolved \nustar\ and \xmm\ pulse profiles in our warped disk model and assume that the pulse profiles are following the changes in strength of the power law and blackbody. 

\subsection{Modeling the Warped Inner Disk}

\cite{hickox2004} found that disk reprocessing is a ubiquitous feature of bright X-ray pulsars. HV05 used a simple warped disk model to describe the differences in shape and phase between the hard and soft pulsations in SMC X-1 as they vary across the superorbital cycle. \cite{hung2010} used the same model to qualitatively describe the pulse profiles in LMC X-4. While these previous works demonstrated the success of the HV05 disk model, neither used observations within a single disk precession cycle to examine the periodicity of the disk and determine if this model can describe pulse behavior over a complete disk cycle. 

We seek to verify whether the HV05 model can reproduce the changes in pulse shape and phase seen over a complete disk precession cycle in both LMC X-4 and SMC X-1. The warped disk model used in this analysis is the same as that presented by HV05 and used by \cite{hung2010}. 

\begin{figure}
\includegraphics[scale=0.61]{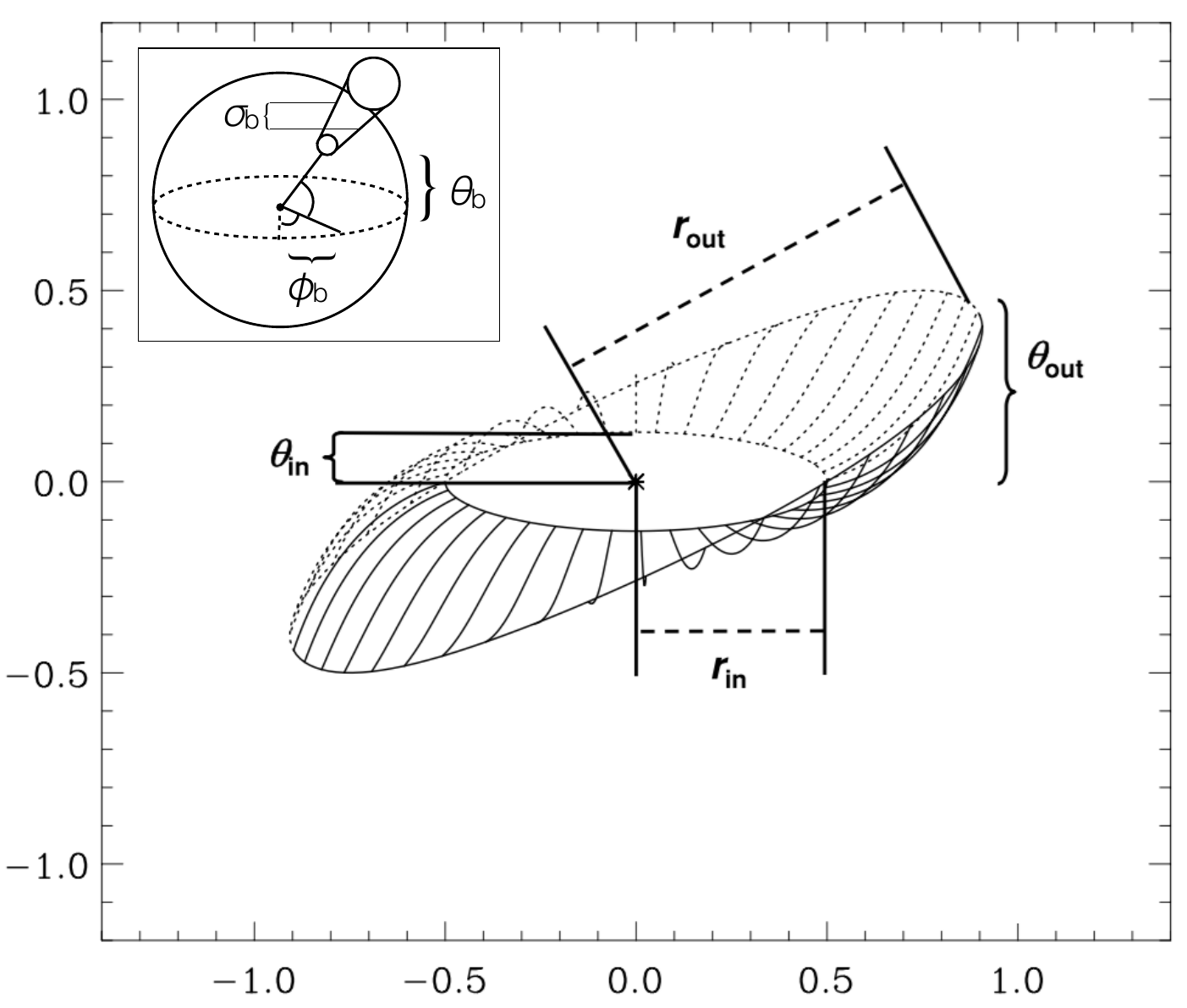}
\caption{A schematic diagram of the disk geometry used in the HV05 model, adapted from Figure 7 in HV05. The inset shows a schematic diagram of the neutron star beam geometry, where a single beam is shown for clarity.}
\label{fig:modelschem}
\end{figure}

HV05 describes the warped inner region of the accretion disk as a series of concentric circles that are inclined and rotated relative to each other. This geometry is based on the well-constrained disk of the bright X-ray binary Her X-1 (\citealt{scott2000,leahy2002}). The precise geometry of this disk is set by the radii of the inner and outer circles and their respective inclination angles ($r_{\text{in}}$, $r_{\text{out}}$, $\theta_{\text{in}}$, $\theta_{\text{out}}$). We provide a schematic diagram of the HV05 disk and beam geometries in Figure \ref{fig:modelschem}.

We also use two simple beam geometries, a pencil and a fan beam, to model the neutron star's beam geometry. In both cases, the beams are modeled as two-dimensional Gaussians with width $\sigma_{\text{b}}$. The location of the beam on the neutron star surface is defined by the angle out of the plane of rotation and the azimuthal angle ($\theta_{\text{b}}$, $\phi_{\text{b}}$). We define the coordinate system such that the poles align with the rotation axis and $\theta = 0$ lies along the equator, and the neutron star's rotation is parallel to the disk axis. In the fan beam model, we also define a beam opening angle ($\theta_{\text{fan}}$), which is set to 0 in the pencil model. For simplicity, the fan beam model is a pure fan beam, without an embedded pencil beam.

In this model, the observer is set at a fixed angle ($\theta_{\text{obs}}$) which, if the neutron star rotates within its orbital plane, is related to the inclination angle for the system by $i = 90$\degree $-\theta_{\text{obs}}$. The beam pattern is then rotated and the regions of the disk visible from the neutron star are illuminated. The disk is assumed to be opaque and it immediately reradiates the absorbed emission as a blackbody spectrum. This assumption requires that the light crossing time and disk cooling time be shorter than the neutron star pulse period. The light crossing time for a disk surface at approximately 10$^{8}$ cm is $\sim$ 10 ms. For the cooling time, \cite{endo2000} suggested that this timescale can be estimated as the thermal energy of the disk divided by the luminosity. For general parameters such as a Compton thick disk, a blackbody temperature of $kT_{\text{BB}}=0.18$ keV, and a soft X-ray luminosity of 10$^{37}$ erg s$^{-1}$, HV05 estimate the cooling time as $\sim10^{-5}$ s. Both the light crossing time and the cooling time are shorter than the pulse periods of LMC X-4 and SMC X-1, and therefore we assume immediate reprocessing by the disk.

Emission seen by the observer is calculated at 30 pulse phases and 8 equally spaced disk phase intervals, where disk phase zero is defined as when the neutron star first emerges from behind the disk, consistent with the start of the superorbital high state. For each beam geometry and disk rotation phase the luminosity of the beam and the luminosity of the disk regions visible to the observer are calculated and simulated hard (beam) and soft (disk) pulse profiles are made. The HV05 model does not include  the effects of light bending on the emission viewed by the observer.

In our model we constrain the disk surface between an inner radius of 0.8 $\times 10^{8}$ cm and an outer radius of 1 $\times 10^{8}$ cm, which HV05 found reproduced the observed SMC X-1 black body temperature. We initially set the observer angle to 20\degree because this agrees with orbital inclination estimates of $\sim$ 70\degree\ for both SMC X-1 and LMC X-4 (\citealt{reynolds1993,vandermeer2007}). While this value worked well for the SMC X-1 models, we found that we could not reproduce the Observation L3 pulse profiles with an observer angle of 20\degree. We tested a range of observer angles from 5--40\degree and found that the Observation L3 pulse profiles could only be reproduced with an observer angle of 40\degree. We set the outer disk angle to be 45\degree\ for both sources; this is within the disk inclination range of 25\degree--58\degree\ estimated for SMC X-1 by \cite{lutovinov2004}, and we found this angle necessary to reproduce the observed LMC X-4 pulse profiles. Our outer disk angle also agrees with hydrodynamic simulations that \cite{larwood1996} used to find stable precession in tilted accretion disks with outer disk angles of 45\degree. We fixed the inner disk angle to a smaller value of 10\degree. We found that a beam half-width of 30\degree\ fit all observations well. While the disk geometry was allowed to vary between LMC X-4 and SMC X-1, we used the same disk parameters to describe the observations from each source, but allowed the beam parameters to change between observations.

When fitting the pulse profiles to data, we allow the overall intensity of the simulated pulses to vary so that the intensity matches that of the observed hard pulsations.

\subsection{Disk Models Output} 

To simulate pulse profiles for both the pencil and fan beam models of LMC X-4, we began fitting pulse profiles with the brightest observation in the data set: Observation L3. We first simulate the hard pulse profile shape to match the observed data and then adjusted the disk parameters until we found reasonable agreement in the soft pulse profiles. We then kept the disk parameters the same for the other three LMC X-4 observations and varied the beam height and azimuth ($\theta_{\text{b}}$, $\phi_{\text{b}}$), which was necessary to match the other pulse shapes in the observation series. We note that these changes in beam location do not necessarily represent physical changes in the accretion column, but rather reflect the varying effects of light bending or other phenomena not included in the HV05 model. For each observation, we allowed the disk to precess and calculate pulse profiles for each precession phase. We fit the three SMC X-1 observations in the same way. The best fit parameters for both the pencil and fan beam configurations are listed in Table \ref{tab:diskpar}.

\begin{figure*}
\centering
\includegraphics[scale=1.2]{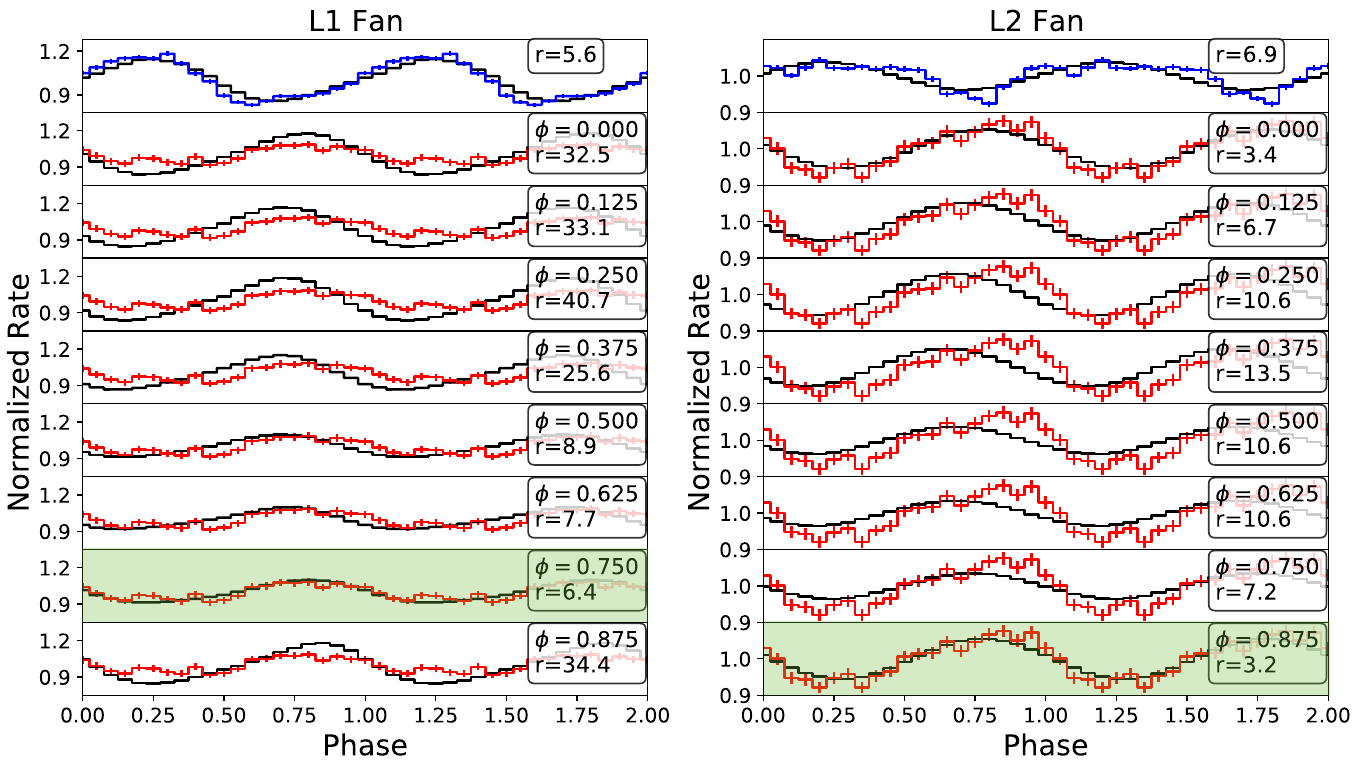}
\includegraphics[scale=1.2]{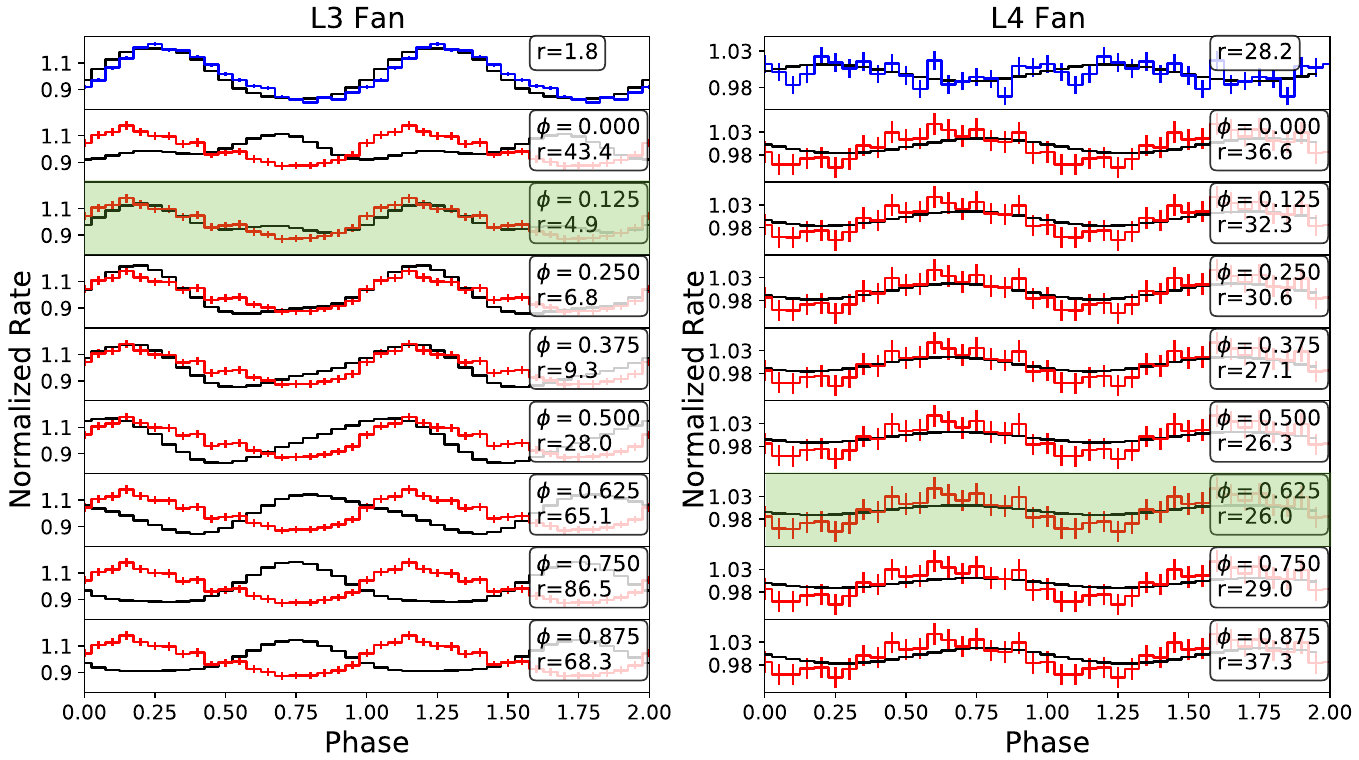}
\caption{Observed hard (blue) and soft (red) pulse profiles compared with simulated (black) pulse profiles from the HV05 fan beam model for the four LMC X-4 observations. For the soft pulses, the simulated pulse profiles from the eight modeled disk rotation phases are shown to demonstrate the effect of disk rotation on pulse shape and phase. The best fit disk rotation phases are highlighted in green.}
\label{fig:ldisksimpp}
\end{figure*}

\begin{figure*}
\centering
\includegraphics[scale=0.65]{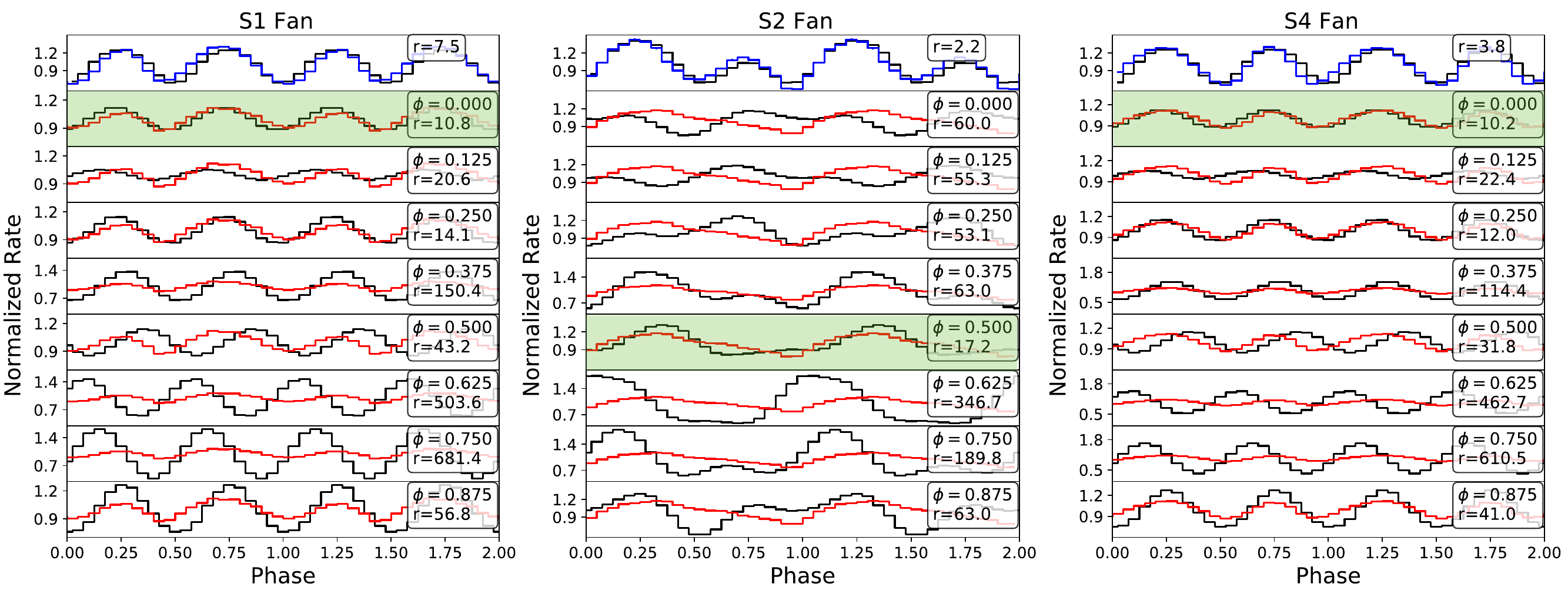}
\caption{Same as Figure \ref{fig:ldisksimpp} for SMC X-1. The best fit disk rotation phases are consistent with a complete disk rotation.}
\label{fig:sdisksimpp}
\end{figure*}

To find the best fit to the soft pulses, we estimated the goodness of fit between the simulated pulse profiles produced at different precession phases and the observed pulse profile by calculating $r=\sum(P_{\text{obs}}(\phi_{\text{spin}}) - P_{\text{sim}}(\phi_{\text{spin}})) / \overline{P_{\text{obs}}}$, where $P_{\text{obs}}$ is the observed pulse profile and $P_{\text{sim}}$ is the simulated pulse profile, and identifying the disk phases with the lowest $r$ value. These best fit disk phases are highlighted in green in Figures \ref{fig:ldisksimpp} and \ref{fig:sdisksimpp}.

In LMC X-4, the HV05 model is able to describe the shape of the hard pulsations with the exception of Observation L4, which has extremely weak pulsations. The lack of pulsations in this observation is possibly due to pulsation dropout in association with super-Eddington accretion flares, and the timing properties of this observation are discussed in \cite{brumback2018b}. For the purposes of this analysis, the effect of weak pulsations in Observation L4 results in poor constraints on the beam profile.

By allowing the disk to precess, the HV05 model successfully reproduces the shape of most of the LMC X-4 soft pulsations in at least one disk phase. However, for the soft pulsations in Observation L3 (which are nearly in phase with the observed hard pulsations), the HV05 model struggles to reproduce the phase of the soft pulsations. This is most likely because of the broad beam parameters necessary to create single peaked pulse profiles. We indicate the best fit disk phases for the fan beam configuration in Figure \ref{fig:ldisksimpp}, however we note that this is likely not a valid constraint on the disk precession phase. The pencil beam configuration produced similar results to those shown in Figure \ref{fig:ldisksimpp}, and so we do not include these figures for the sake of conciseness.

We also find good fits to the observed hard pulse profiles for SMC X-1. We show the results of the fan beam configuration in Figure \ref{fig:sdisksimpp}, and again do not show the similar results from the pencil beam configuration for the sake of space. We found that the HV05 model struggled to reproduce the soft pulses observed in Observation S2. The challenges in simulating these pulse profiles likely arise from the hard pulsations having a double peaked profile; when the hard profile was double peaked the model strongly preferred a soft profile that was double peaked as well. We found that for a double peaked hard profile, the HV05 model was not able to return a single peaked soft profile as broad as the observed profile.

\begin{figure*}
\centering
\begin{tabular}{c}
	\includegraphics[scale=0.7]{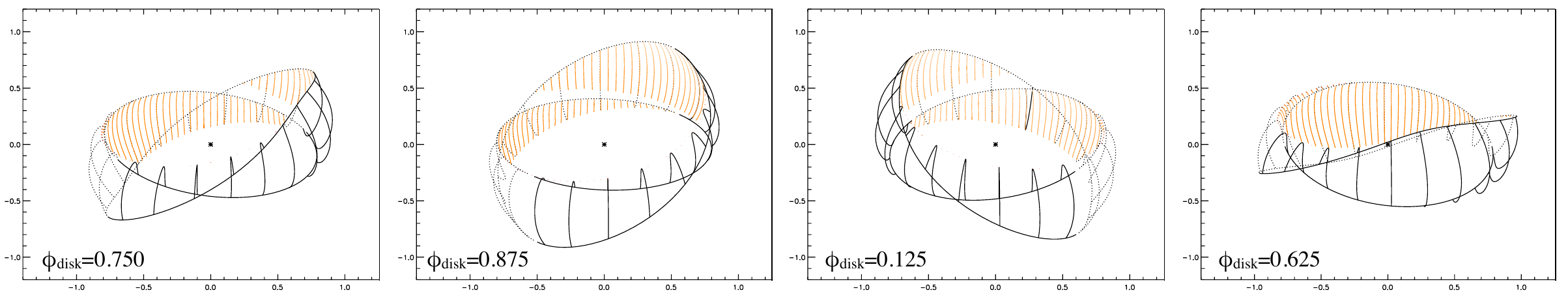} 
\end{tabular}
\caption{Disk models for the four best fit disk precession phases from the LMC X-4 Observations, showing the possible disk geometry of a complete disk precession cycle. Units are 10$^{8}$ cm.}
\label{fig:ldiskim}
\end{figure*}

\begin{figure*}
\centering
\begin{tabular}{c}
	\includegraphics[scale=0.65]{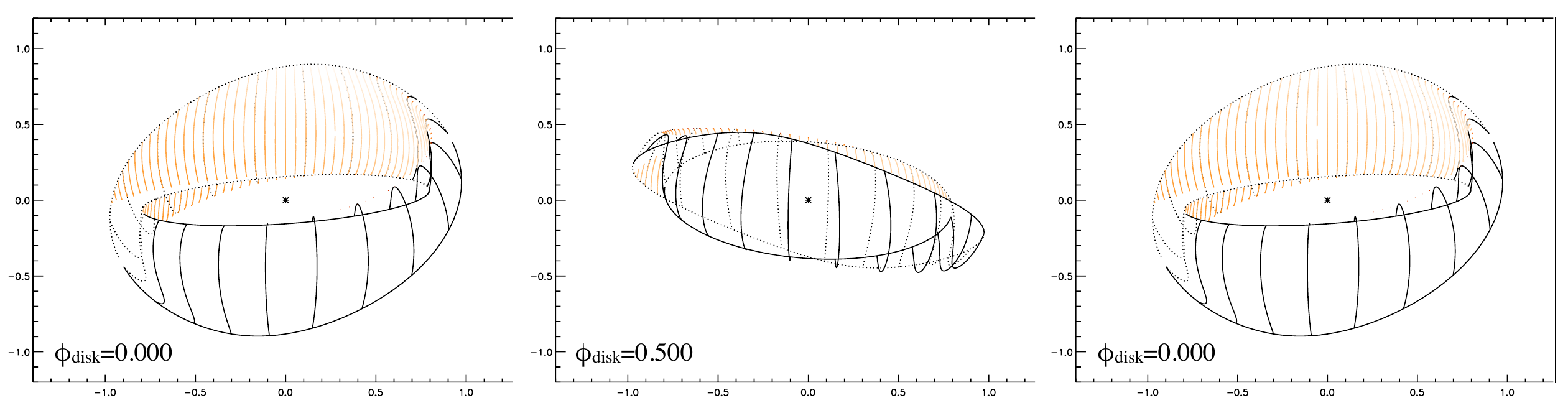} 
\end{tabular}
\caption{Disk models for the three best fit disk precession phases from the SMC X-1 Observations, showing the possible disk geometry of a complete disk precession cycle. Units are 10$^{8}$ cm.}
\label{fig:sdiskim}
\end{figure*}

Despite modeling challenges presented by Observations L3 and S2, we find that in both LMC X-4 and SMC X-1 the disk phase values corresponding to our best fit soft pulse profiles are consistent with a complete precession cycle of the inner accretion disk (Figures \ref{fig:ldiskim} and \ref{fig:sdiskim}).

Our simulation of the hard and soft profiles confirmed general conclusions made by \cite{hickoxvrtilek2005}, including that the pulse profile shape is more dependent on the beam geometry than the disk geometry, and that the double and single peaked pulse profiles seen in these sources strongly prefer non-antipodal beam geometry. This preference can be seen in Table \ref{tab:diskpar}, where negative values of $\theta_{\text{b}1}$ and $\theta_{\text{b}2}$ are only found in Observation L3 and Observation S2.

\section{Discussion} \label{sec:disc}

Changes in pulse shape as a function of superorbital phase have been previously examined in LMC X-4 and SMC X-1 by \cite{hickoxvrtilek2005}, \cite{neilsen2004}, and \cite{hung2010}. These works, and several of the references therein, strongly imply that the relative changes between hard and soft pulse profiles is caused by reprocessed emission from a warped, precessing inner disk. However, none of these previous analysis of these two X-ray binaries include broadband X-ray coverage over a single superorbital cycle. The joint campaigns carried out by \xmm\ and \nustar\ that are presented in this work represent the first sampling of a complete superorbital cycle with full hard X-ray coverage in these sources.

The joint observations of LMC X-4 and SMC X-1 were carried out so that the first and fourth observation in each series occurred at the same superorbital phase. In these observations, we would expect to see similarities in spectral shape and pulse shape. The results of our spectral and timing analyses for LMC X-4 and SMC X-1 confirm these expectations: in SMC X-1 the pulse profiles and phase-averaged spectra of Observations S1 and S4 are consistent. In LMC X-4 the results are complicated by instances of pulse dropout within Observations L1 and L4. This resulted in different spectral shapes between these two observations and significantly weaker pulsations in Observation L4. However, even though the Observation L4 pulsations are weak, the hard and soft pulses are approximately 180\degree\ out of phase, which is also seen in Observation L1.

The HV05 warped disk model offers an opportunity to simulate pulse profiles for a simplified source geometry of either a pencil or fan beam that irradiates a warped inner disk. We found that both pencil and fan beam geometries can reproduce observed pulse profiles in LMC X-4 and SMC X-1, and thus we cannot conclusively say that one beam geometry is preferred over another. Regardless of beam geometry, we found that pulse profiles where the hard and soft pulsations were out of phase generally preferred non-antipolar beam geometries. This geometry preference agrees with those found by HV05.

By allowing the inner disk to precess in the HV05 model, we found disk precession phases which best described the phase of soft pulses relative to the hard ones. For both LMC X-4 and SMC X-1, our best fit disk precession phases for each observation indicate a smoothly rotating disk. We also find that the disk precession phases are periodic with superorbital phase, meaning the first and last observation in each series shows the disk returning to its approximate initial position. The success of this model further confirms that disk precession can reproduce the observed changes in pulse profiles.

In order to fit the observed pulse profiles, especially of LMC X-4 which vary significantly in relative phase and strength, it was necessary to allow the beam geometry to change between observations. Rather than suggesting that the beam parameters change significantly with superorbital phase, these changes likely represent the effects of varying height in the accretion column, which in turn influence the light bending or other relativistic phenomena that are not included in the HV05 model. One way to further constrain the beam geometries within the HV05 model is to use a source with a highly constrained accretion disk geometry. The ideal target for such further analysis is Hercules X-1, whose 35 day superorbital cycle has been modeled by \cite{leahy2002}. With firm constraints on the disk geometry inputs to the HV05 model, we could possibly see whether a pencil or fan beam geometry is preferred, and compare the results to those presented in this work (\citealt{brumbackprep}).

The geometries included in the HV05 warped disk model are quite simple and likely not a complete representation of the complexities of the inner accretion flow and accretion column structure (e.g.\ \citealt{miyasaka2013}). Future work could update the beam geometries with more complex beam structures (e.g.\ \citealt{koliopanos2018, iwakiri2019}) or physically motivated accretion column models (e.g.\ \citealt{sokolovaprep}) and include the effects of light bending (e.g.\ \citealt{falknersuba}; \citealt{falknersubb}). Despite the simplified nature of the HV05 model, the success of the HV05 model suggests that tomography is a viable method of probing the structure of magnetized accretion flows in neutron star binaries, which can be difficult to resolve observationally. Constraining the warped disk and beam geometry in pulsars with superorbital modulation can shed light on interactions between the accretion disk and the pulsar magnetosphere.

\section{Summary} \label{sec:summary}

In this work we perform a broad band spectral and timing analysis of the X-ray binaries LMC X-4 and SMC X-1 within a single superorbital period. Both of these sources display superorbital periods that are attributed to warped precessing inner accretion disks. We observed each source jointly with \xmm\ and \nustar\ at four epochs during a single superorbital cycle, and found that the pulse profiles and phase-averaged spectra display the periodicity expected from sources with precessing inner disks. We also apply the HV05 warped disk model and find that these observed changes in pulse profiles can be modeled by reflection off of a simple precessing disk. Modeling the geometry of the inner disk and neutron star beam offers a way to observationally examine magnetic accretion flows around neutron stars.

\acknowledgements
We would like to thank the anonymous referee for their comments, which improved the manuscript. We would like to thank the \textit{NuSTAR} Galactic Binaries Science Team and G. Vasilopoulos for comments and contributions. MCB acknowledges support from NASA grant numbers NNX15AV32G and NNX15AH79H. This research made use of NuSTARDAS, developed by ASDC (Italy) and Caltech (USA), XMM-SAS developed by ESA, ISIS functions (ISISscripts) provided by 
ECAP/Remeis observatory and MIT (http://www.sternwarte.uni-erlangen.de/isis/), and the MAXI data provided by RIKEN, JAXA and the MAXI team.

\pagebreak
\bibliography{my_bib}

\end{document}